\documentclass[traditabstract]{aa} 

\usepackage{graphicx}
\usepackage{float}
\usepackage{txfonts}
\usepackage{longtable}
\usepackage{lscape}
\usepackage{hyperref}
\usepackage{soul}
\hypersetup{colorlinks=true, citecolor=blue}
\newcommand{\hi}{{\sc H\,i}}
\newcommand{\hii}{{\sc H\,ii}}
\newcommand{\nii}{{\sc N\,ii}}
\newcommand{\sii}{{\sc S\,ii}}

\begin{document}

   \title{A Virgo Environmental Survey Tracing Ionised Gas Emission (VESTIGE)\\ XXI. Statistical properties of individual \hii\ regions in perturbed galaxies
      }
   \subtitle{}
  \author{A. Boselli\inst{1,**}, 
 	  M. Fossati\inst{2,3},
	  Y. Roehlly\inst{1},
	  M. Boquien\inst{4}, 
	  J. Braine\inst{5},
          P. C{\^o}t{\'e}\inst{6},
    	  J.C. Cuillandre\inst{7},
	  B. Epinat\inst{8,1},
    	  L. Ferrarese\inst{6},
    	  S. Gwyn\inst{6},
	  G. Hensler\inst{9}
       }

\institute{     
                Aix Marseille Univ, CNRS, CNES, LAM, Marseille, France\thanks{Scientific associate INAF - Osservatorio Astronomico di Cagliari, Via della Scienza 5, 09047 Selargius (CA), Italy}
                \email{alessandro.boselli@lam.fr}
        \and
         	Universit\'a di Milano-Bicocca, piazza della scienza 3, 20100 Milano, Italy
	\and
		INAF - Osservatorio Astronomico di Brera, via Brera 28, 20121 Milano, Italy
	\and
        	Laboratoire d'Astrophysique de Bordeaux, Univ. Bordeaux, CNRS, B18N, all\'ee Geoffroy Saint-Hilaire, 33615 Pessac, France	
 	\and
		Universit\'e C\^ote d'Azur, Observatoire de la C\^ote d'Azur, CNRS, Laboratoire Lagrange, 06000, Nice, France
	\and
        	National Research Council of Canada, Herzberg Astronomy and Astrophysics, 5071 West Saanich Road, Victoria, BC, V9E 2E7, Canada
	\and
       	 	AIM, CEA, CNRS, Universit\'e Paris-Saclay, Universit\'e Paris Diderot, Sorbonne Paris Cit\'e, Observatoire de Paris, PSL University, F-91191 Gif-sur-Yvette Cedex, France
	\and
		French-Chilean Laboratory for Astronomy, IRL 3386, CNRS and Universidad de Concepci\'on, Departamento de Astronomia, Barrio Universitario s/n ,Concepti\'on, Chile
    	\and
        	Department of Astrophysics, University of Vienna, T\"urkenschanzstrasse 17, 1180 Vienna, Austria
                 }

\authorrunning{Boselli et al.}
\titlerunning{VESTIGE: statistical properties of \hii\ regions in perturbed galaxies}

   \date{}

  \abstract  
{We use narrow-band H$\alpha$+[\nii ] imaging data gathered during the Virgo Environmental Survey Tracing Ionised Gas Emission (VESTIGE), a blind survey of the Virgo cluster carried 
out with MegaCam at the Canada-French-Hawaii telescope (CFHT), to identify \hii\ regions in 385 galaxies showing ionised gas emission. After excluding objects where the emission is not associated
to star formation and edge-on systems, we identify 76\,645 \hii\ regions in 322 star-forming galaxies and study their physical properties for those above the completeness limit of the 
survey ($L(H\alpha$) $\geq$ 10$^{37}$ erg~s$^{-1}$, 34\,358 regions). The present work is focused on perturbed cluster galaxies, identified as those having a reduced amount of atomic hydrogen 
when compared to similar objects in the field. We derive composite luminosity functions, diameter and electron density distributions, and several scaling
relations, and compare them to those already derived for gas-rich, unperturbed systems identified during the VESTIGE survey.
The analysis shows that the statistical and physical properties of \hi\ gas-deficient cluster galaxies are different from those of unperturbed systems, with perturbed objects having a 
steeper faint-end slope and a brighter characteristic H$\alpha$ luminosity than gas-rich galaxies. The difference in the two distributions comes principally from the outer
disc (outside the effective radius). Perturbed and unperturbed systems share a similar \hii\ size distribution, while gas-poor objects hosts higher electron density regions
than \hi\--rich systems. The analysis of the scaling relations indicates that perturbed objects have, on average, a lower number of \hii\ regions per unit stellar mass and disc surface 
than unperturbed systems, with differences increasing with the \hi\--deficiency parameter, principally in the outer disc where \hii\ regions are less present in gas-poor systems. 
This systematic difference is also observed in the H$\alpha$ luminosity of the first ranked and first three ranked \hii\ regions, which is reduced in \hi\--deficient 
systems with respect to gas-rich objects. All these differences can be explained 
in the framework of galaxy evolution in rich environments, where their hydrodynamic interaction with the surrounding intracluster medium
(ram pressure) removes the gas outside-in quenching the star formation activity in the outer disc once the atomic hydrogen is removed.
}


   \maketitle
%

\section{Introduction}

Galaxies evolving in rich environments such as clusters and groups are affected by different kind of perturbations which are able to modify their star 
formation activity. These perturbing mechanisms can be broadly divided into two main families (see Boselli \& Gavazzi 2006, 2014 for a review): 
gravitational perturbations with other galaxies and/or with the gravitational potential well of the high-density region itself (tidal interactions, 
harassment, tidal stirring, Merritt 1983, Moore et al. 1996) and hydrodynamic interactions with the hot ($T_{ICM}$ $\simeq$ 10$^7$-10$^8$ K) and 
diffuse ($\rho_{ICM}$ $\simeq$ 10$^{-3}$ cm$^{-3}$; Sarazin 1986) intracluster medium (ICM) permeating the high density region (ram pressure stripping (RPS), 
thermal evaporation, starvation; Gunn \& Gott 1972; Cowie \& Songaila 1977; Larson et al. 1980). All these mechanisms are able to affect the cold gas 
content and distribution of the perturbed galaxies, indirectly modifying on different timescales their activity of star formation (e.g. Boselli et al. 
2022b). It is indeed known that gravitational perturbations remove gas from the outer disc and at the same time favor gas infall in the inner regions 
through disc instabilities and bar formation, thus feeding starburst activity in the nucleus (Ellison et al. 2008, 2011). In a starvation scenario, 
when the infall of fresh gas on the stellar disc is stopped once galaxies enter the cluster halo (Larson et al. 1980), star formation is uniformly 
reduced at all galactocentric distances (Boselli et al. 2006). On the contrary, ram pressure removes the gas outside-in, producing truncated gas 
and star-forming discs (e.g. Koopman \& Kenney 2004a,b, 2006; Boselli et al. 2022b). 

While the overall effects of the different kind of interactions on the global and large scale star formation activity of perturbed galaxies is 
now well understood, it is still not clear which are the effects on the scale of individual \hii\ regions ($\simeq$ 100 pc). It is known from 
observations and simulations (Fujita \& Nagashima 1999; Bekki \& Couch 2003; Nehlig et al. 2016; Steyrleithner et al. 2020; Troncoso-Iribarren 
et al. 2020; Boselli et al. 2021; Liz\'ee et al. 2021; Zhu et al. 2024) that the gas can be locally compressed when galaxies are interacting 
with their surrounding environment. Indeed, it has been observed that at the front edge of a RPS episode, at the interface 
between the interstellar medium (ISM) of the perturbed galaxy and the surrounding hot ICM, the gas can be compressed increasing its local density. In these regions, 
strong episodes of star formation have been observed, with the formation of giant \hii\ regions (e.g. CGCG 97-73, Gavazzi et al. 1995, 2001; 
NGC 4654, Vollmer 2003; ESO 137-001, Fossati et al. 2016;
IC 3476, Boselli et al. 2021, J201, Bellhouse et al. 2019). It is, however, totally unknown whether external perturbations 
have statistically significant effects on the physical properties of \hii\ regions, i.e. whether there are evident effects on their physical properties
such as their luminosity, size, and electron gas density. 

VESTIGE is a blind narrow-band (NB) H$\alpha$+[\nii ] imaging survey of 
the Virgo cluster carried out with MegaCam at the CFHT. The survey, which covers the entire Virgo 
cluster up to its virial radius ($\simeq$ 104$^{\circ 2}$), detected 385 galaxies with a clear H$\alpha$ emission.
Given that this ionising radiation comes directly from young and massive stars, the NB H$\alpha$ emission is a direct tracer of the 
recent activity of star formation (Kennicutt 1998; Boselli et al. 2009). The VESTIGE data, which have been gathered during exceptional 
and uniform seeing conditions (seeing $\simeq$ 0.73\arcsec\, corresponding to $\simeq$ 60 pc angular resolution at the mean distance of the 
cluster of 16.5 Mpc), are providing us with a unique sample of galaxies spanning a wide range in morphological type and stellar mass to study the 
effects of external perturbations on the physical properties of individual \hii\ regions on a statistically significant sample of objects. 
For this purpose, we identified thanks to the use of the \textsc{HIIphot} (Thilker et al. 2000) $\simeq$ 80\,000 \hii\ regions and studied 
their properties in a subsample of unperturbed systems (Boselli et al. 2025). The purpose of this work is that of comparing the physical 
and statistical properties of \hii\ regions in perturbed galaxies to those already derived for the subsample of gas-rich, unperturbed objects 
using a set of uniform data to minimise any possible selection bias. In Sec. 2 we describe the sample, in Sec. 3 the data and we analyse 
them in Sec. 4. Discussion and conclusions are given in Sec. 5. We also present the full dataset in dedicated Appendices.

\section{Sample}

The sample analysed in this work includes all the galaxies with H$\alpha$ emission detected during the VESTIGE survey
where individual \hii\ regions of diameter $\gtrsim$ 60 pc can be easily resolved and their photometric parameters 
correctly measured (accuracy $\lesssim$ 10-20\%, Boselli et al. 2025). The
accurate definition of the sample is given in Boselli et al. (2025). To summarise, 
we exclude from the H$\alpha$ detected sources a few objects where the H$\alpha$ emission is diffuse and not associated 
to star-forming regions, such as in M87 (Boselli et al. 2019) or in some lenticular galaxies (Boselli et al. 2022a).
The identification of \hii\ regions is particularly challenging in highly inclined systems. We thus exclude edge-on galaxies, 
where the axis ratio is $b/a$ $<$ 0.25 ($a$ and $b$ are the major and minor axes measured on the $i$-band NGVS image),
which roughly corresponds to discs with inclinations $i$ $\gtrsim$ 75 degrees. We also excluded those \hii\ regions
located outside the stellar disc of the parent galaxy measured at the 25.5 mag~arcsec$^{-2}$ $B$-band isophotal diameter (Binggeli et al. 1985). 
Finally, to grant the completeness of the identified \hii\ regions, 
we limit the analysis to those galaxies hosting \hii\ regions with $L(H\alpha)$ $\geq$ 10$^{37}$ erg~s$^{-1}$. With these
limitations, the final sample is composed of 322 star-forming galaxies spanning a wide range in morphological type
(from massive spirals and lenticulars to Magellanic irregulars, blue compact dwarfs (BCD), and dwarf ellipticals)
and stellar mass (10$^7$ $\lesssim$ $M_{star}$ $\lesssim$ 10$^{11}$ M$_{\odot}$). Galaxies are assumed at the mean distance 
of the cluster substructure to which they belong, with distances according to Cantiello et al. (2024): 16.5 Mpc for galaxies 
belonging to cluster A (M87) and to the low velocity cloud (LVC), 15.8 Mpc for those belonging to cluster B (M49) 
and cluster C (M60), 23 Mpc in the W$^{\prime}$ cloud, and 32 Mpc for clouds W and M (see Fig. 3 in Boselli et al. 2023b for details).

\begin{figure*}
\centering
\includegraphics[width=0.33\textwidth]{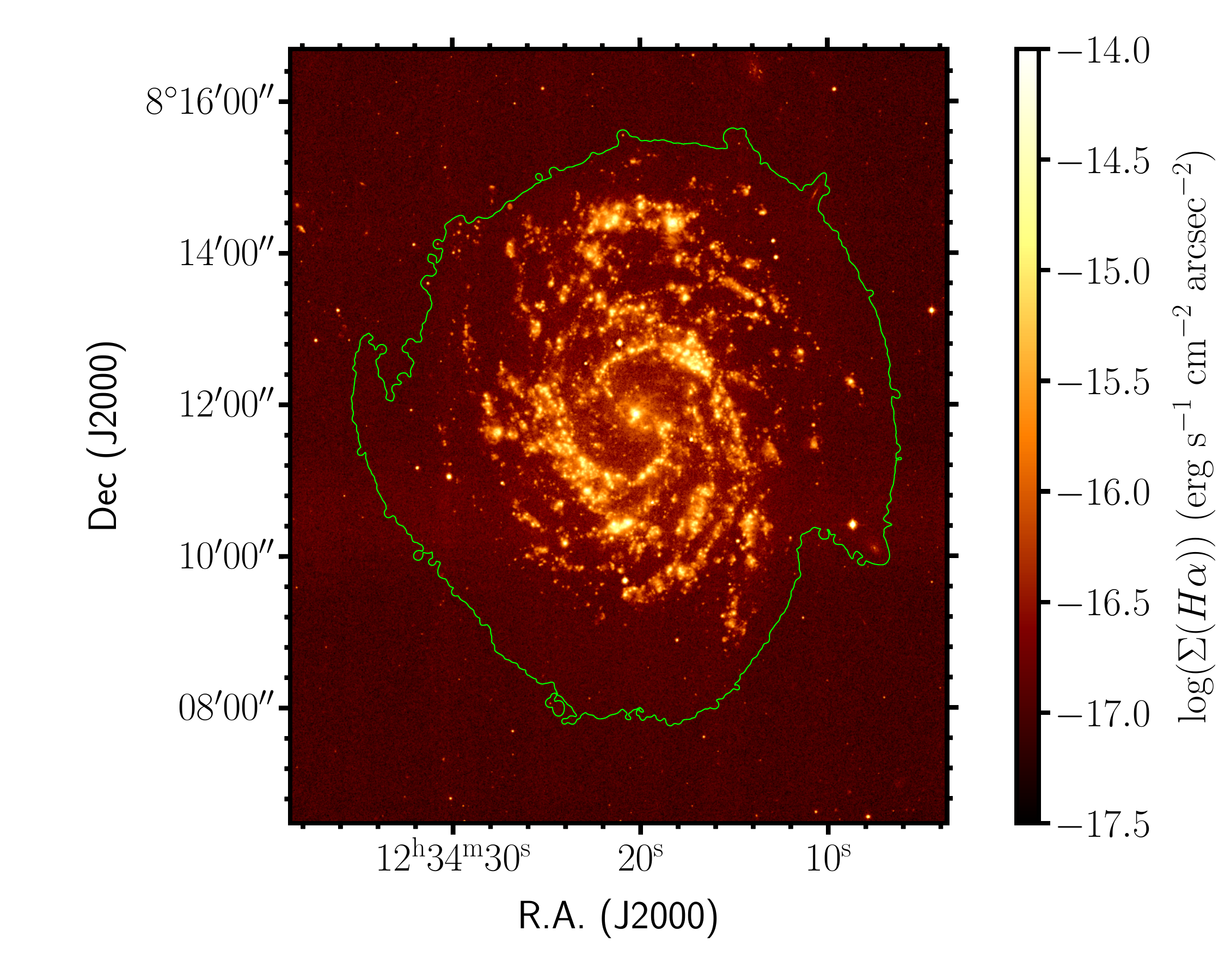}
\includegraphics[width=0.33\textwidth]{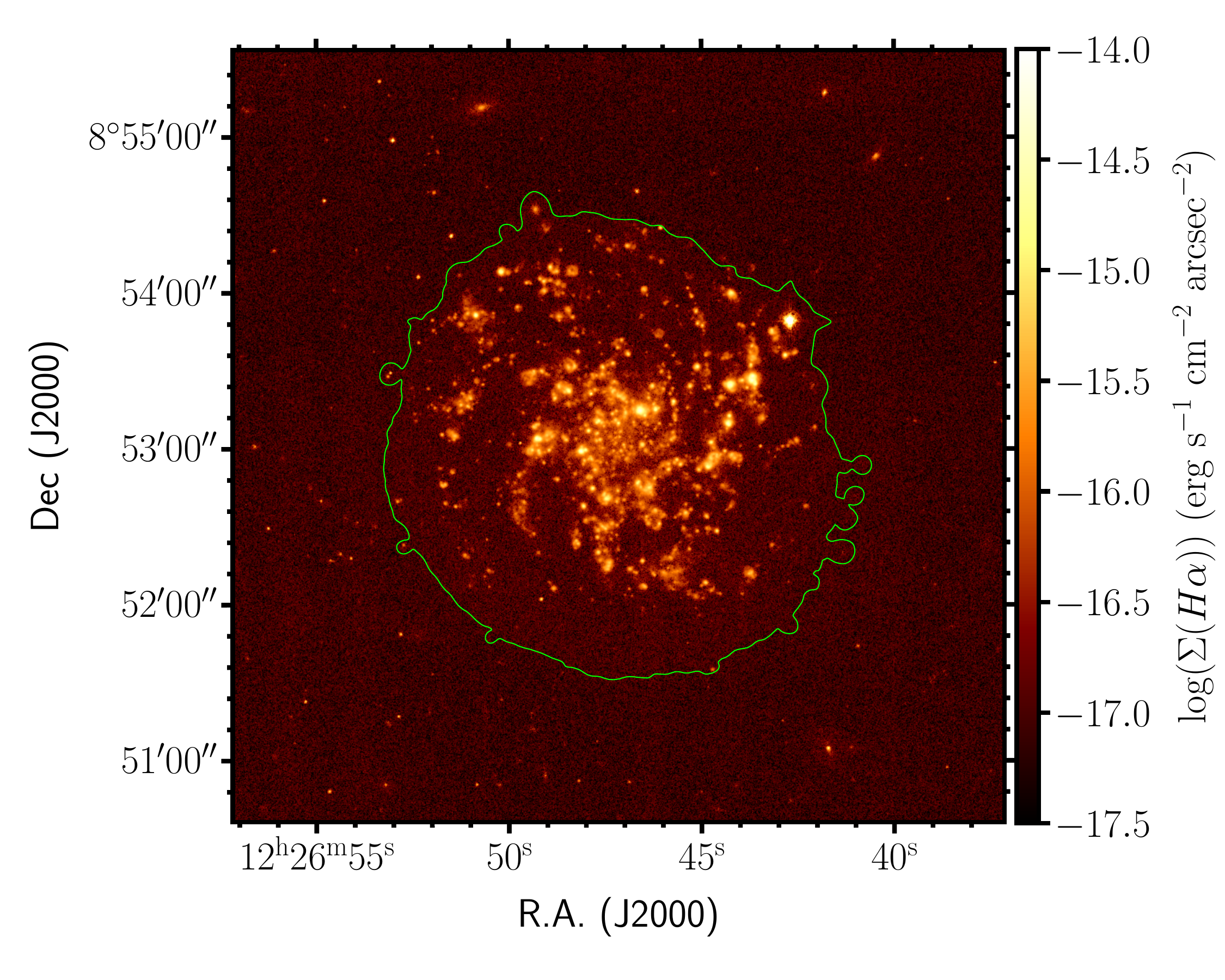}
\includegraphics[width=0.33\textwidth]{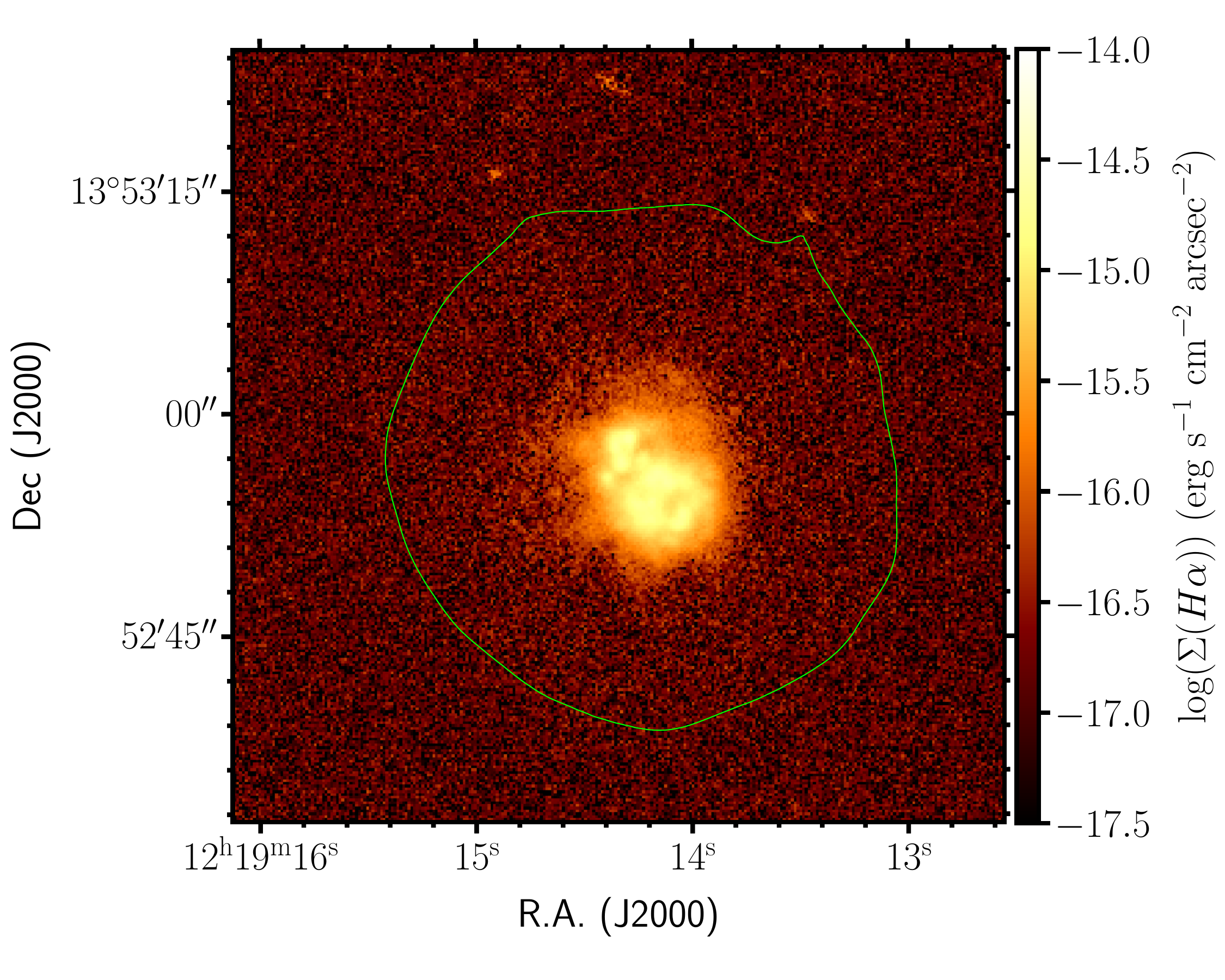}\\
\includegraphics[width=0.33\textwidth]{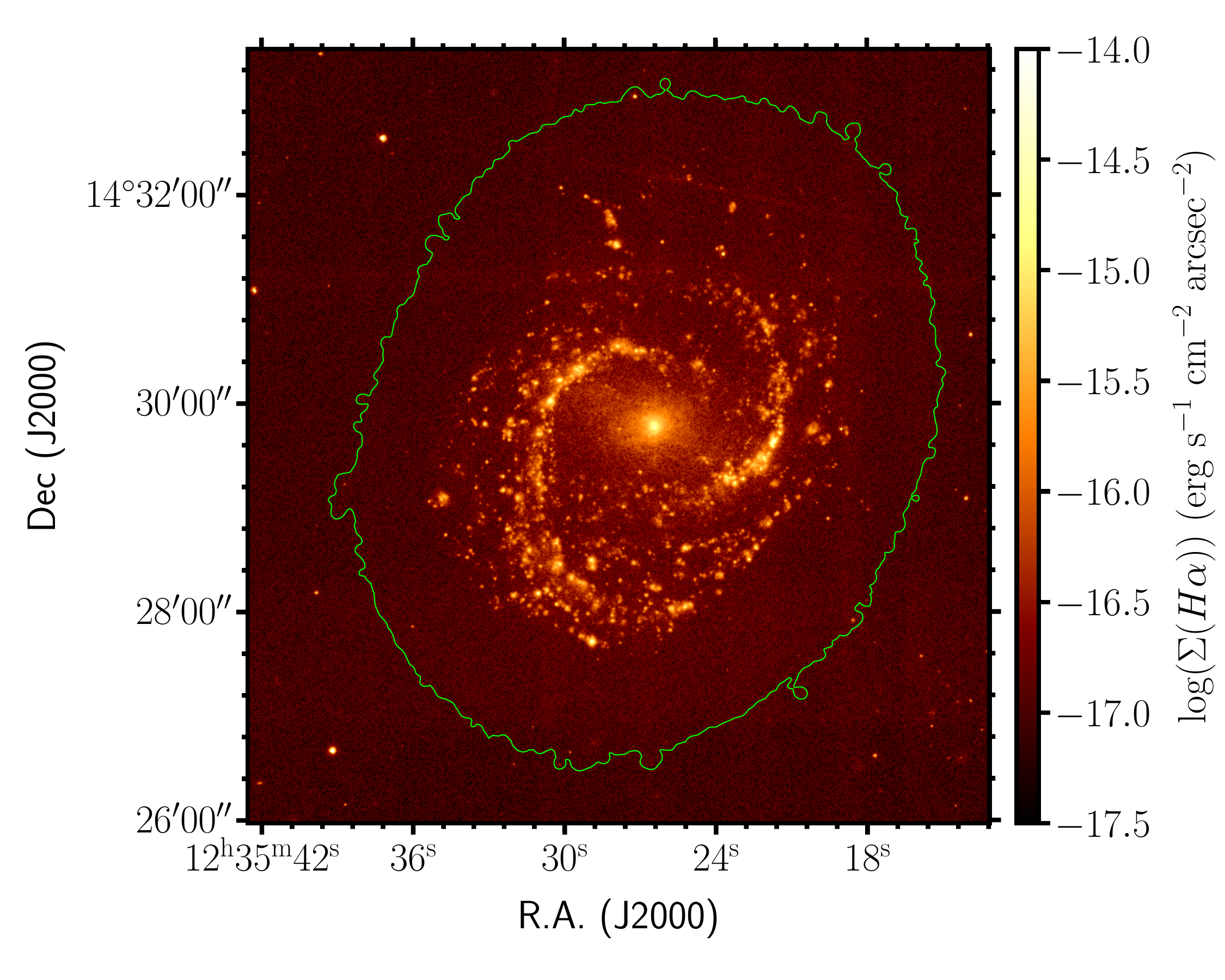}
\includegraphics[width=0.33\textwidth]{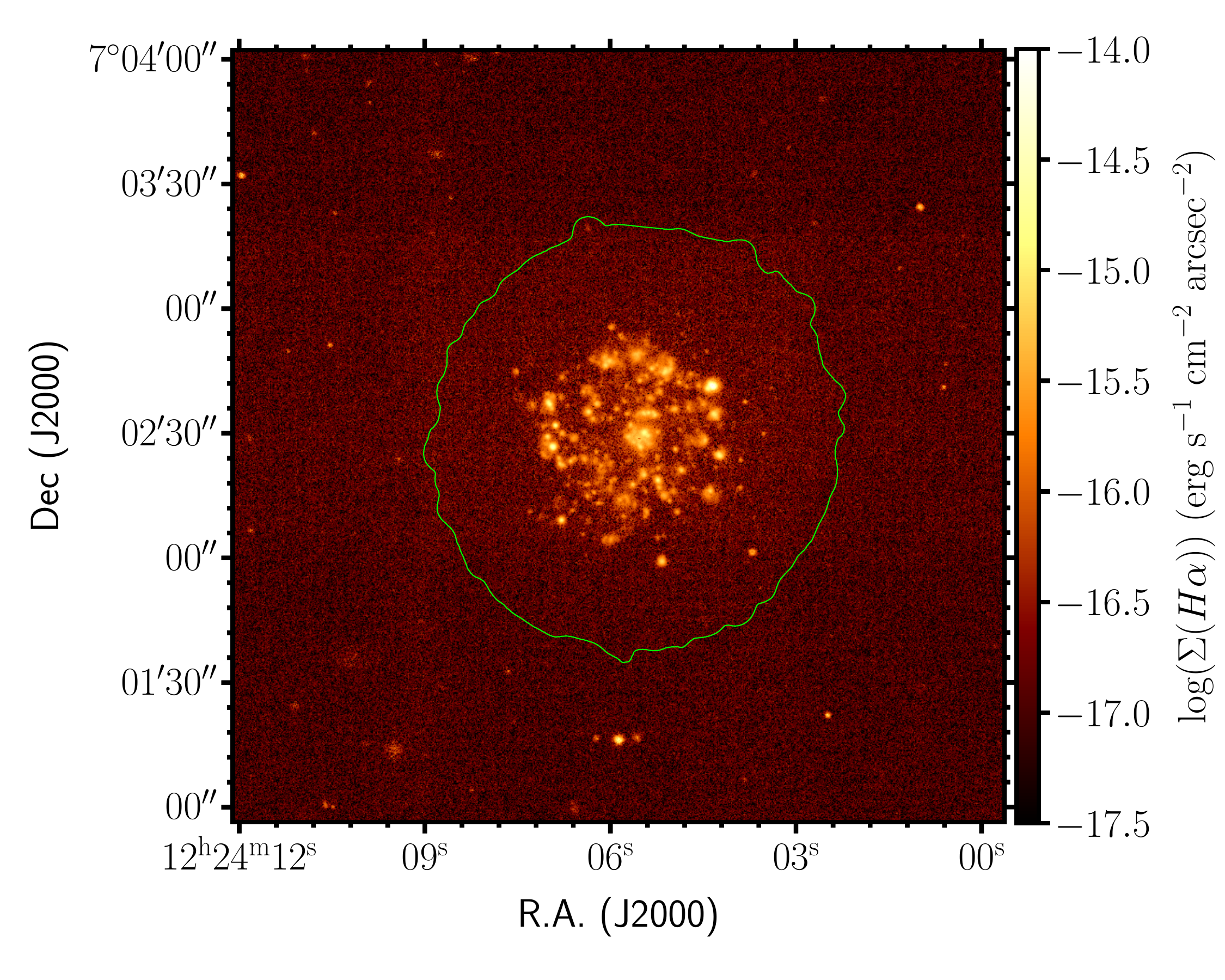}
\includegraphics[width=0.33\textwidth]{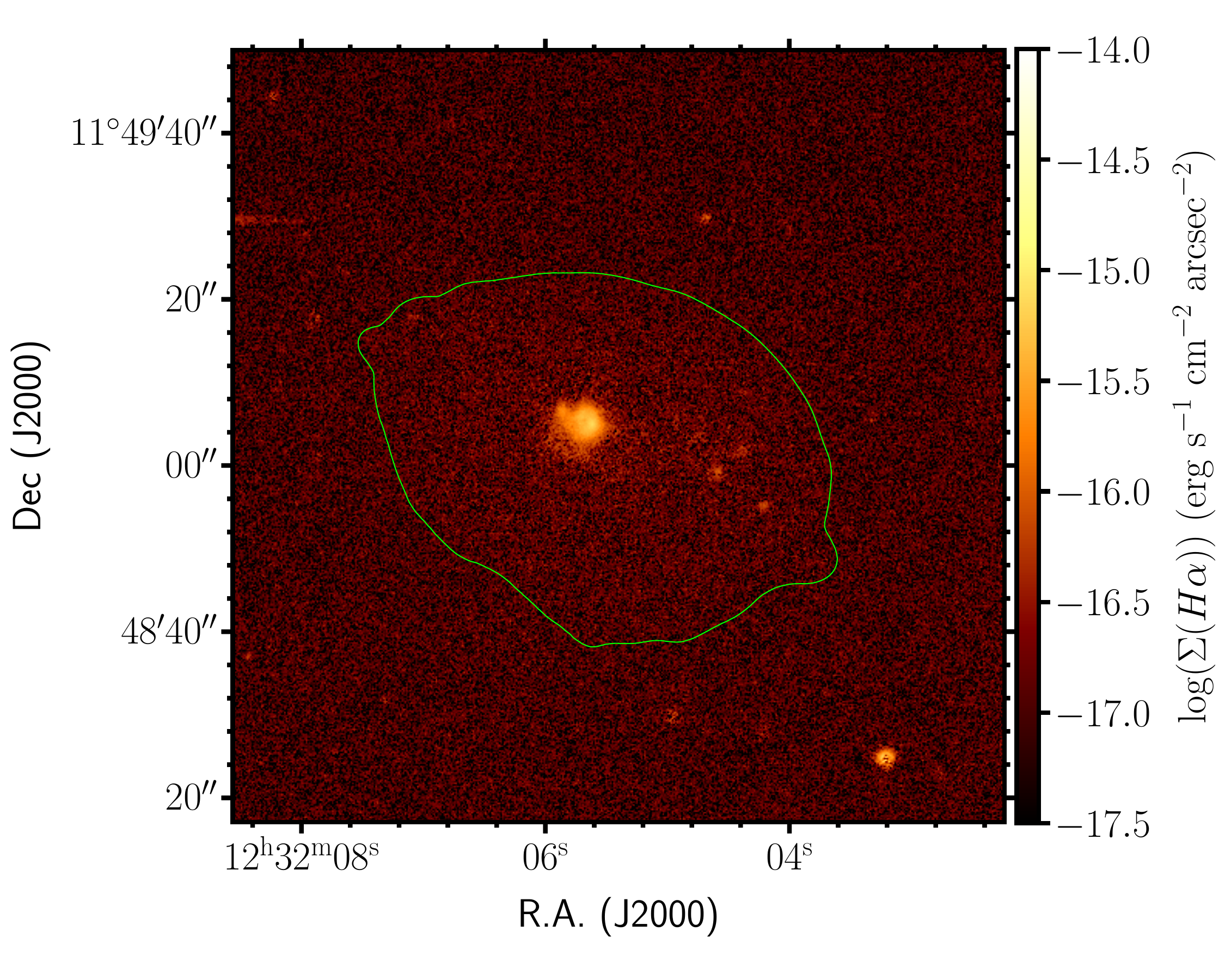}\\
\caption{Continuum-subtracted H$\alpha$ images of representative galaxies in the unperturbed (upper row) 
and perturbed (lower row) samples in different bins of stellar mass: $M_{star}$ $>$ 10$^{10}$ M$_{\odot}$ 
(upper left column: NGC4535-VCC1555, lower left column: NGC4548-VCC1615),
10$^9$ $<$ $M_{star}$ $\leq$ 10$^{10}$ M$_{\odot}$ (upper central column: NGC4411b-VCC939, lower central column: IC3267-VCC697), 
10$^8$ $<$ $M_{star}$ $\leq$ 10$^{9}$ M$_{\odot}$ (upper right column: VCC334; lower right column: IC3466-VCC1411).
The green contour shows the $r$-band 25.0 mag arcsec$^{-2}$ isophote indicating the extension of the stellar disc.
}
\label{Haimage}%
\end{figure*}

To identify perturbed galaxies, we use the \hi\--deficiency parameter defined as the logarithmic difference between the 
expected and the observed \hi\ mass of the target galaxies (Haynes \& Giovanelli 1984), where the expected gas mass is derived for each object 
using the calibration of Cattorini et al. (2023). This parameter has been proven to statistically and quantitatively 
probe the degree of interaction that a galaxy is suffering in a rich environment (e.g. Boselli \& Gavazzi 2006; Cortese et al. 2021; 
Boselli et al. 2022b). The perturbed sample, which 
is the main target of this work, is composed of 258 galaxies with $HI-def$$>$0.4. We use here the 64 galaxies of the 
unperturbed sample ($HI-def$$\leq$0.4), extensively analysed in Boselli et al. (2025), as reference to statistically 
quantify the effects of the interactions on the star-forming properties of the perturbed sample. Figure \ref{Haimage}
show some representative examples of continuum-subtracted H$\alpha$ images of \hi-normal and \hi-deficient galaxies in the two samples in 
different bins of stellar mass. In perturbed galaxies, \hii\ regions are mainly located in the inner disc and are generally less luminous
than those observed in unperturbed systems of similar stellar mass.

\begin{figure}
\centering
\includegraphics[width=0.49\textwidth]{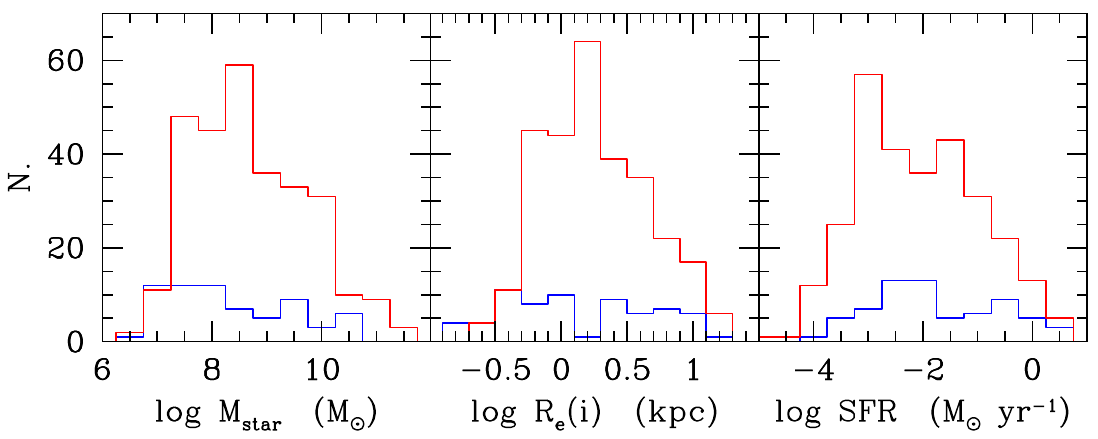}\\
\caption{Comparison of the distributions of the stellar mass (left panel), $i$-band effective radius (central panel), and star formation rate
(right panel) of the unperturbed (blue filled histogram) and perturbed (red empty histogram) samples. }
\label{mstardist}%
\end{figure}

Table 1 of Boselli et al. (2025) gives the fraction of perturbed and unperturbed objects in different bins of stellar mass.
Figure \ref{mstardist} shows that the stellar mass, $i$-band effective radii, and star formation rate distributions of the 
two samples are covering the same dynamic range, with fairly comparable mean values ($<$log$M_{star}$ $>$ = 8.42$\pm$1.13 M$_{\odot}$, 
$<$log$R_e(i)$ $>$ = 0.14$\pm$0.55 kpc, $<$log$SFR$ $>$ = -1.68$\pm$1.10 M$_{\odot}$ yr$^{-1}$ for the unperturbed sample, and 
$<$log$M_{star}$ $>$ = 8.76$\pm$0.98 $_{\odot}$, 
$<$log$R_e(i)$ $>$ = 0.28$\pm$0.39 kpc, $<$log$SFR$ $>$ = -1.95$\pm$1.02 M$_{\odot}$ yr$^{-1}$ for the perturbed sample\footnote{ 
A Kolmogorov-Smirnov test gives $p$-values = 0.01, 0.21, 0.001 that the two populations are drawn by the same distribution.}). 
We thus do not expect any major systematic effect 
in the following analysis related to differences in the physical properties of the two samples.

\section{Narrow-band H$\alpha$ imaging data}

The data analysed in this work have been gathered during the VESTIGE H$\alpha$ narrow-band imaging survey of the Virgo cluster.
The details of the observing strategy, data acquisition and reduction is given 
in Boselli et al. (2018a), while the description of the multifrequency data used in the analysis, the identification of the \hii\ 
regions and the derivation of their physical parameters, as well as the estimate of the completeness of the survey are reported 
in Boselli et al. (2025). We refer the interested readers to these works, while we give here only a very brief summary.

VESTIGE is an untargeted survey of the Virgo cluster up to its virial radius ($r_{200}$ = 1.55 Mpc, Ferrarese et al. 2012,
corresponding to 104$^{\circ 2}$). The observations have been taken in the NB filter MP9603 centred on the H$\alpha$ line ($\lambda_c$
=6591~\AA ; $\Delta\lambda$=106~\AA ) and in the broad-band $r$, necessary for the subtraction of the stellar continuum.
The NB filter includes the emission of the H$\alpha$ Balmer line ($\lambda$=6563~\AA ) and of the two [{\nii}] 
lines at $\lambda$=6548-6583~\AA \footnote{Hereafter we refer to the H$\alpha$ $+$[{\nii}] band simply as H$\alpha$, unless 
otherwise stated.}, and it has its peak transmissivity ($T$ $\simeq$ 92\%) in the velocity range 
$-$1140 $\leq$ $v_{hel}$ $\leq$ 3700 km s$^{-1}$, which is optimal to sample the velocity dispersion of galaxies within the 
Virgo cluster (Boselli et al. 2014a; 2018a). The observations, which have been gathered using a large dither pattern to 
optimise the determination of the sky background, have been taken with 2~h integration in the NB filter and 12 minutes in the
broad-band $r$ filter. With this integration time, the sensitivity of the survey reaches 
$f(H\alpha)$ $\simeq$ 4$\times$10$^{-17}$ erg~s$^{-1}$~cm$^{-2}$ (5$\sigma$) for point sources, and
$\Sigma(H\alpha)$ $\simeq$ 2$\times$10$^{-18}$ erg~s$^{-1}$~cm$^{-2}$~arcsec$^{-2}$ (1$\sigma$ after smoothing the data to 
$\simeq$3\arcsec\ angular resolution), with photometric uncertainties of $\leq$0.02-0.03 mag in both bands.
We recall that at this sensitivity for point sources we are able to detect \hii\ regions of luminosity 
$L(H\alpha)$ $\geq$ 1.3$\times$10$^{36}$ erg~s$^{-1}$ at the typical distance of the cluster (16.5 Mpc), 
which is lower than the H$\alpha$ luminosity expected 
for the emission of a single O star and comparable to that of a single early-B star (Sternberg et al. 2003).
We these data, we are thus potentially able to detect all the ionising emission within the star-forming regions 
of the observed galaxies. 

The subtraction of the stellar continuum is secured using a combination of the broad-band $r$ and $g$ filters, this last available
thanks to the NGVS survey (Ferrarese et al. 2012). The identification of the \hii\ regions is done using the 
\textsc{HIIphot} data reduction pipeline (Thilker et al. 2000). The spectacular angular resolution of the data, which have a 
mean seeing of $FWHM_{PSF}$ = 0.73\arcsec, allows us to resolve \hii\ regions down to scales of $\simeq$ 60 pc (we recall 
that at the mean distance of the cluster 1\arcsec\ = 80 pc). The comparison with integral field unit (IFU) spectroscopic 
obtained with MUSE for a few objects in common confirms that the accuracy in the flux derivation is within $\lesssim$ 10\%. 
The data are corrected for dust attenuation and [{\nii}] contamination using a variety of spectroscopic data or simple scaling 
relations whenever these are not available, as described in Boselli et al. (2023a). As in previous works 
(Boselli et al. 2021, 2025), equivalent diameters (defined as the diameter of the circle of area corresponding to that 
of the observed \hii\ region) are corrected for the effects of the point-spread function (PFS) following Helmboldt et al. (2005):

\begin{equation}
{D_{eq} = D_{HII}\frac{\sqrt{FWHM^2_{mod} - FWHM^2_{PSF}}}{FWHM_{mod}}},
\end{equation}

\noindent
where $D_{HII}$ is the output diameter of \textsc{HIIphot}, $FWHM_{mod}$ = $\sqrt{FWHM_{maj}FWHM_{min}}$ is the effective circular FWHM from
the Gaussian model fit of the 2D line intensity emission, and $FWHM_{PSF}$ is the seeing as obtained from fitting bright stars in the 
images with Gaussian models. 
Finally, we derived electron densities $n_e$ of individual \hii\ regions following 
Scoville et al. (2001) adopting the relation (case B recombination, Osterbrock \& Ferland 2006):

\begin{equation}
n_\mathrm{e} = 43\bigg[\frac{(L_{cor}(H\alpha)/10^{37} ~\rm{erg~s^{-1}})(T/10^4 ~\rm{K})^{0.91}}{(D_{eq}/10 ~\rm{pc})^3}\bigg]^{1/2} ~~[\rm{cm^{-3}}],
\end{equation} 

\noindent
where $L_{cor}(H\alpha)$ is the H$\alpha$ luminosity of the individual \hii\ regions corrected for [{\nii}] contamination and dust attenuation, 
and $T$ the gas temperature (here assumed to be $T$ = 10\,000 K). To avoid large uncertainties in the adopted corrections,
equivalent diameters and electron densities 
are only derived for those regions where the correction for the effect of the PSF is less than 50\%. 

Finally, for completeness we limit the following analysis to those \hii\ regions with H$\alpha$ luminosity corrected for [{\nii}] contamination and dust attenuation  
$L(H\alpha)$ $\geq$ 10$^{37}$ erg~s$^{-1}$ (see Boselli et al. 2025).
To minimise seeing-related effects, we limit, unless specified, the analysis of size-related entities to those 
regions where the correction is less than 50\%\ . 
With these criteria, the perturbed sample of 258 galaxies has 49\,315 \hii\ regions, out of which 21\,080 with a H$\alpha$ luminosity 
corrected for dust attenuation and [{\nii}] contamination $L(H\alpha)$ $\geq$ 10$^{37}$ erg~s$^{-1}$. Of these, 22\,041 and 10\,781 have an estimate of the effective
diameter derived using a correction $\leq$ 50\%, respectively. These numbers are compared to those of the unperturbed sample 
analysed in Boselli et al. (2025) in Table \ref{Tabstat}. Worth noticing is the fact that the mean number of \hii\ regions of $L(H\alpha)$ $\geq$ 10$^{37}$ erg~s$^{-1}$ per galaxy
is 191 in the perturbed sample vs. 427 in the unperturbed sample.

\begin{table}
\caption{Number of \hii\ regions in the perturbed and unperturbed samples
}
\label{Tabstat}
{
\[
\begin{tabular}{ccc}
\hline
\noalign{\smallskip}
\hline
Sample								& Unperturbed		&  Perturbed \\
								& N.			& N.		\\
\hline
All								& 27\,330		& 49\,315	\\
$L(H\alpha)$ $\geq$10$^{37}$					& 13\,278		& 21\,080 \\
$D_{eq}/D_{HII}$ $\geq$ 50\% 					& 11\,293		& 22\,041	\\
$D_{eq}/D_{HII}$ $\geq$ 50\% \& $L(H\alpha)$ $\geq$10$^{37}$	& 6\,520		& 10\,781	\\
\hline
\end{tabular}
\]
}
\end{table}

\section{Analysis}

The purpose of this work is to analyse the statistical properties of \hii\ regions in perturbed galaxies and compare them
to those observed in the reference sample of unperturbed objects analysed in Boselli et al. (2025). Being observed during the same VESTIGE survey,
and located at the same distance, the comparison of the two samples minimises any systematic distance-related bias in the analysis. Indeed, any possible selection effect
in the derivation of the H$\alpha$ luminosity and size of individual \hii\ regions should affect in a similar way the data of gas-rich and gas-poor systems. 
For this purpose, we follow the same 
structure presented in Boselli et al. (2025), i.e. we first derive the statistical properties of the sample through the derivation of the
H$\alpha$ luminosity function of \hii\ regions, we then measure their diameter and electron density distributions, the luminosity-diameter relation, 
and finally present the major scaling relations where systematic effects between the two samples of perturbed and unperturbed galaxies have been observed.

\subsection{Integrated distributions}

\subsubsection{Composite luminosity function}

\begin{figure}
\centering
\includegraphics[width=0.49\textwidth]{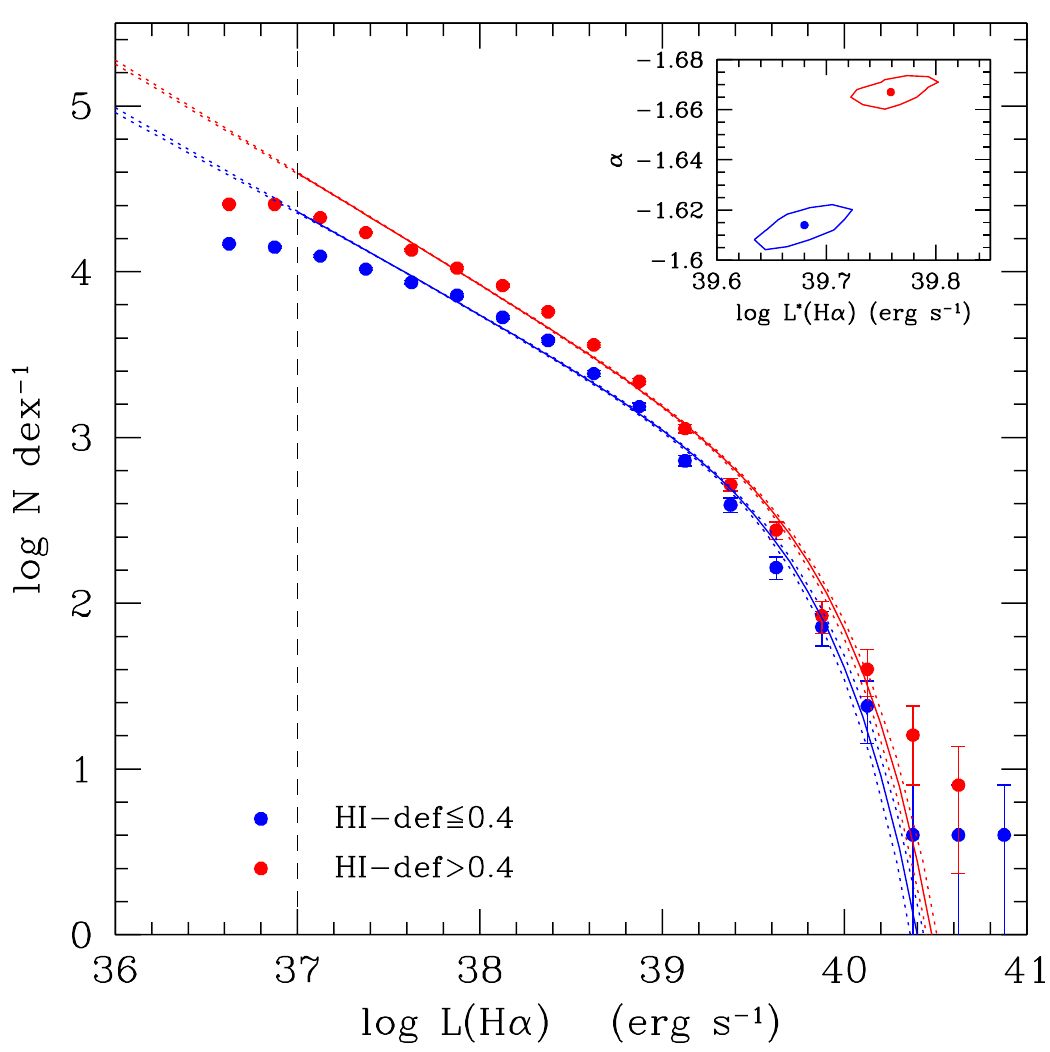}\\
\caption{Composite luminosity function of the {\hii} regions detected by \textsc{HIIphot} on the selected galaxies for unperturbed
($HI-def$ $\leq$ 0.4; blue dots) and perturbed ($HI-def$ $>$ 0.4; red dots) galaxies. The H$\alpha$ luminosities of 
individual {\hii} regions have been corrected for dust attenuation and [{\nii}] contamination. The solid and dotted lines
indicate the best-fit and 1$\sigma$ confidence regions for the Schechter luminosity function parametrisation. The vertical dashed line 
shows the completeness of the survey. The small panel in the top right corner indicates the 1$\sigma$ probability distribution of the 
fitted Schechter function parameters.}
\label{LFHIIall}%
\end{figure}

\begin{figure}
\centering
\includegraphics[width=0.49\textwidth]{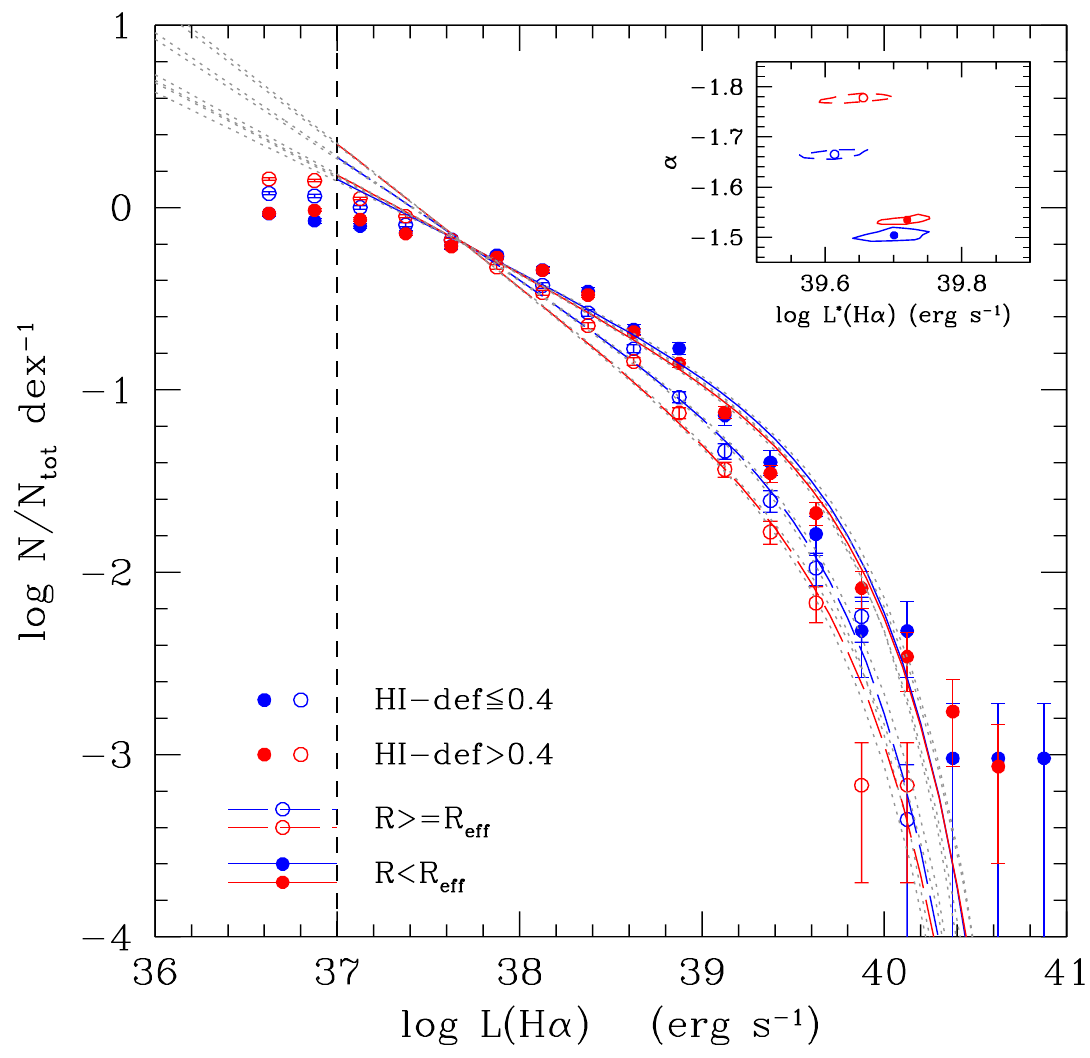}\\
\caption{Composite luminosity function of the {\hii} regions detected within (filled dots, solid lines) and outside (empty dots, dashed lines) 
the $i$-band effective radius of the target galaxies normalised to the total number of {\hii} regions with $L(H\alpha)$ $\geq$ 10$^{37}$ erg~s$^{-1}$
within ($N_{tot}$=4\,194; 9\,292) and outside ($N_{tot}$=9\,084; 11\,788) the effective radius for unperturbed ($HI-def$ $\leq$ 0.4; blue dots and lines)
and perturbed ($HI-def$ $>$ 0.4; red dots and lines) galaxies, respectively. The vertical dashed line 
shows the completeness of the survey. The small panel in the top right corner indicates the 1$\sigma$ probability distribution of the 
fitted Schechter function parameters.}
\label{LFHIIReffallnew}%
\end{figure}

Figure \ref{LFHIIall} shows the composite H$\alpha$ luminosity functions derived by counting the number of \hii\ regions per bin of H$\alpha$ 
luminosity (0.25 dex in log scale) in perturbed and unperturbed systems. We fit the two distributions with a Schechter (1976) function 
following the same procedure proposed by Metha et al. (2015) and Fossati et al. (2021) as successfully done in Boselli et al. (2025). For this purpose,  
we define ${\mathcal L}$ = $\log_{10}(L)$ and we fit the distribution with the relation:

\begin{equation}
{\rm{\Phi}({\mathcal L})d{\mathcal L} = \rm{ln(10)\Phi^*10}^{({\mathcal L}-{\mathcal L}^*)(1+\alpha)}\rm{exp}(-10^{({\mathcal L}-{\mathcal L}^*)})d{\mathcal L}},
\end{equation}

\noindent
where $d{\mathcal L}$ = $d\log_{10}(L)$, ${\mathcal L}^*$=$\log_{10}L^*$ (in units of erg~s$^{-1}$), and we derive the posterior 
distribution and the best-fit parameters using the MULTINEST Bayesian algorithm
(Feroz \& Hobson 2008, Feroz et al. 2019). The terms $L^*$, $\Phi^*$, and $\alpha$ are the characteristic luminosity at the knee
of the distribution, the number of objects at $L^*$, and the slope of the distribution at the faint end, respectively.
We recall that, as defined, the MULTINEST Bayesian algorithm used to derive the best-fit 
parameters of the luminosity function does not depend on the binning which is here used only for
a graphical representation of the distribution.
Consistently with Boselli et al. (2025), we limit the fit of the Schechter function to $L(H\alpha)$ $\geq$ 10$^{37}$ erg~s$^{-1}$
where the distribution is complete. The best fits, derived using data corrected for dust attenuation and [{\nii}] contamination, 
reproduce fairly well the two distributions, although they both underestimate
the number of bright \hii\ regions above $L(H\alpha)$ $\gtrsim$ 10$^{40}$ erg~s$^{-1}$ and over-predict the number of \hii\ regions 
below $L(H\alpha)$ $\lesssim$ 10$^{37.5}$ erg~s$^{-1}$. Despite this similar general behaviour, the fitted functions are 
statistically different as indicated by the 1$\sigma$ confidence regions for the Schechter luminosity function parametrisation: 
the composite H$\alpha$ luminosity function of the $perturbed~sample$ has a brighter H$\alpha$ characteristic luminosity
and a steeper faint end slope than the one of the $unperturbed~sample$ (see Fig. \ref{LFHIIall} and Table \ref{TabLF}). 

Figure \ref{LFHIIReffallnew} shows the composite H$\alpha$ luminosity function of \hii\ regions measured within (solid line, filled dots) and outside (dashed line, empty circles) the
$i$-band effective radius $R_{eff}$ of the target galaxies for unperturbed (blue dots and lines) and perturbed (red dots and lines) systems. The comparison of
the fitted Schechter functions and of their derived parameters (see Table \ref{TabLF}) indicates that while the perturbed and unperturbed systems 
have comparable H$\alpha$ distributions within the inner discs, they significantly differ in the outer galaxy regions. Here, the two distributions have a 
similar characteristic H$\alpha$ luminosity, but perturbed systems have a steeper faint end slope ($\alpha$=$-$1.78$\pm$0.01) than unperturbed objects 
($\alpha$=$-$1.66$\pm$0.01). 

\begin{figure*}
\centering
\includegraphics[width=0.49\textwidth]{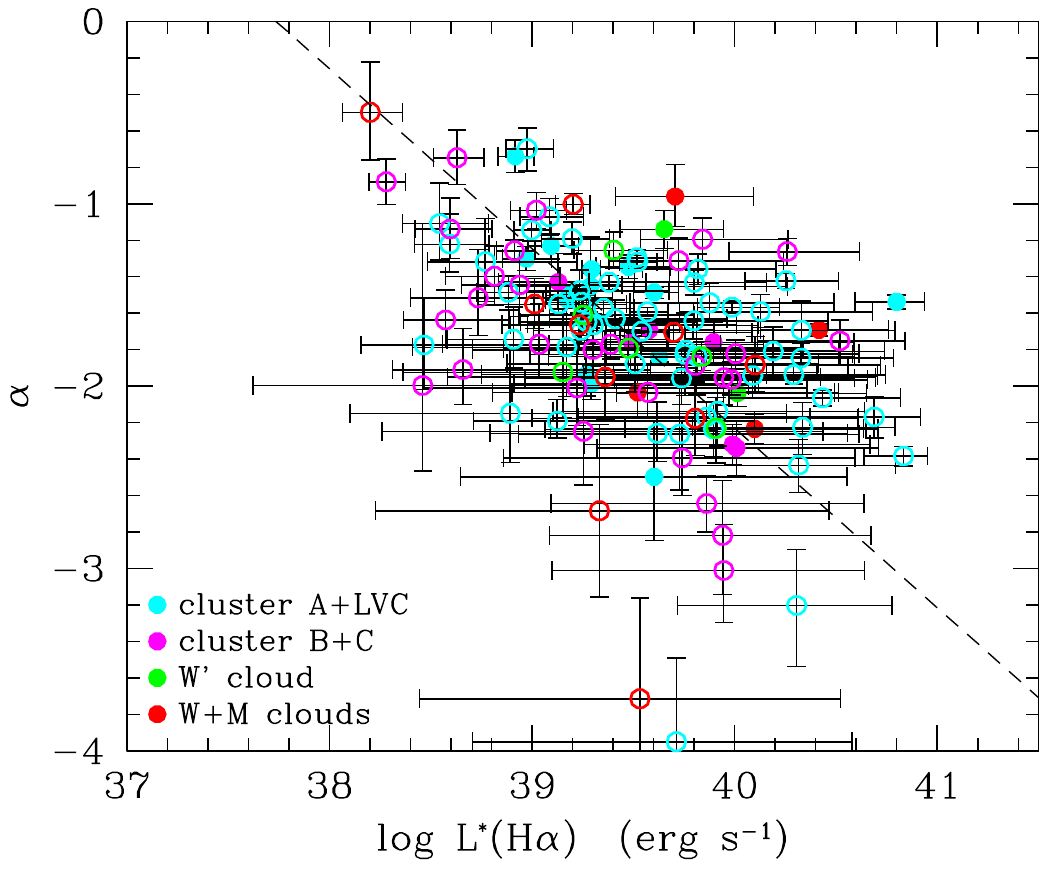}
\includegraphics[width=0.49\textwidth]{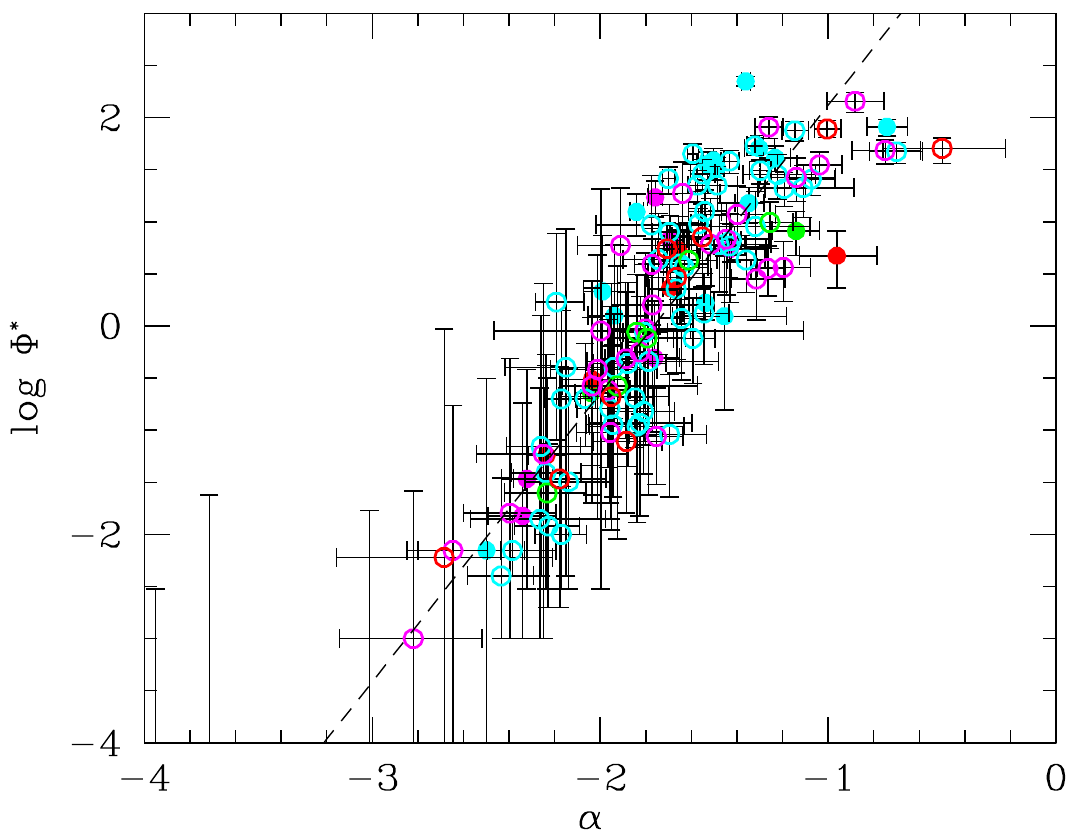}\\
\caption{Left: relation between the faint end slope $\alpha$ of the luminosity function and the characteristic luminosity $L^*(H\alpha)$
of individual galaxies with more than 20 {\hii} regions brighter than $L(H\alpha)$ $\geq$ 10$^{37}$ erg~s$^{-1}$. Right:
relation between the characteristic number of objects $\Phi^*$ and the faint end slope $\alpha$ of the luminosity function.
Different colours are used for galaxies belonging to 
different cluster substructures: cyan for cluster A and LVC, 16.5 Mpc; magenta for cluster B, 15.8 Mpc;
green for W' cloud, 23 Mpc; and red for W and M clouds, 32 Mpc. The H$\alpha$ luminosities of 
individual {\hii} regions are corrected for dust attenuation and [{\nii}] contamination. Filled and open circles are for galaxies
with $HI-def$ $\leq$ 0.4 and $HI-def$ $>$ 0.4, respectively. The black dashed line shows the best fit to the data (bisector fit, Isobe et al. 1990)
measured on the full sample of 127 galaxies. The best fits are
$\alpha$ = $-$0.99($\pm$0.08) $\times$ $\log L^*(H\alpha)$ + 37.18($\pm$1.63), $\rho$=0.49, $\sigma$=0.40, where $\rho$ and $\sigma$ are the Spearman
correlation coefficient and the dispersion of the relation in the left panel, and  
$\log \Phi^*$ = 2.77($\pm$0.07) $\times$ $\alpha$+4.88($\pm$0.11), $\rho$=0.87, $\sigma$=0.21 in the right panel.}
\label{individualparametersall}%
\end{figure*}

The sample includes 127 galaxies with more than 20 \hii\ regions brighter than $L(H\alpha)$ $\geq$ 10$^{37}$ erg~s$^{-1}$, out of which
27 and 100 included in the unperturbed and perturbed sample, respectively. For these galaxies the number of \hii\ regions is sufficiently high 
to allow a fit of the Schechter function on individual objects\footnote{The fit diverges on two galaxies of the perturbed sample.}. 
The mean of the best fit parameters derived for individual galaxies are given in Table \ref{TabLF}. They do not show any systematic difference between
gas-rich and gas-poor systems probably because of the large dispersion in their distribution. 
Figure \ref{individualparametersall} shows the relationship between the best 
fit parameters for the whole sample with symbols coded according to the membership of galaxies to the different cluster substructures, 
located at different distances. Contrary to what is observed in the unperturbed sample (Boselli et al. 2025), probably because of a
low number of objects, we observe a weak but statistically
significant relation between the faint end slope $\alpha$ and the characteristic H$\alpha$ luminosity $L^*(H\alpha)$. We also observe 
a strong relation between $\Phi^*$ and the $\alpha$ parameter.

\begin{table}
\caption{Parameters of the Schechter function.
}
\label{TabLF}
{
\[
\resizebox{\columnwidth}{!}{\begin{tabular}{ccccc}
\hline
\noalign{\smallskip}
\hline
Sample					& N.objects		&  $\alpha$     & log$L^*(H\alpha)$     & $\Phi^*$ \\
					&			&		& erg~s$^{-1}$		&		\\
\hline
Unperturbed				&			&		&			&		\\
\hline
Composite				& 13\,278		& $-$1.61$\pm$0.01& 39.68$\pm$0.04	& 226$\pm$21	\\
Mean$^a$				& 27			& $-$1.66$\pm$0.43& 39.57$\pm$0.44	& -		\\
$R$ $\geq$ $R_{eff}$			& 9\,084		& $-$1.66$\pm$0.01& 39.61$\pm$0.05	& 136$\pm$17	\\
$R$ $<$ $R_{eff}$			& 4\,194		& $-$1.50$\pm$0.02& 39.70$\pm$0.06	& 113$\pm$14	\\
\hline
Perturbed				&			&		&			&		\\
\hline
Composite				& 21\,080		& $-$1.67$\pm$0.01& 39.76$\pm$0.04	& 247$\pm$23	\\
Mean$^a$				& 100			& $-$1.78$\pm$0.55& 39.50$\pm$0.57	& -		\\
$R$ $\geq$ $R_{eff}$			& 11\,788		& $-$1.78$\pm$0.01& 39.66$\pm$0.06	& 97$\pm$15	\\
$R$ $<$ $R_{eff}$			& 9\,292		& $-$1.53$\pm$0.01& 39.72$\pm$0.04	& 214$\pm$18	\\
\hline

\end{tabular}}
\]
Notes: All fits are done for {\hii} regions with $L(H\alpha)\geq10^{37}$ erg~s$^{-1}$. \\
$a$): Mean values derived for galaxies having more than 20 individual {\hii} regions.
Uncertainties given here are the dispersion in the parameter distribution. 
}
\end{table}

\subsubsection{Composite diameter distribution}

\begin{figure}
\centering
\includegraphics[width=0.49\textwidth]{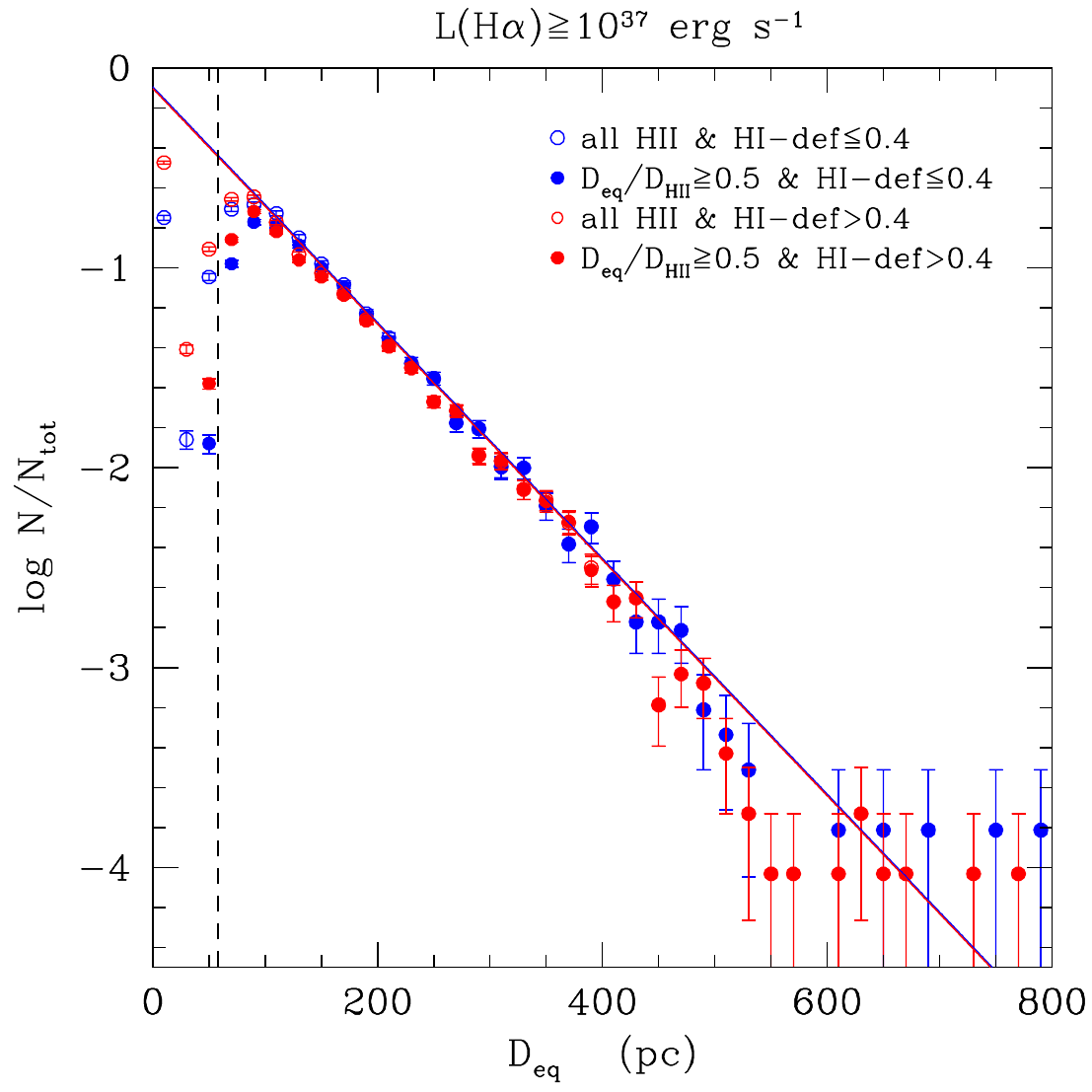}\\
\caption{Normalised distribution of the observed (empty dots) equivalent diameters 
for {\hii} regions with $L(H\alpha)$ $\geq$ 10$^{37}$ erg~s$^{-1}$ for unperturbed ($HI-def$ $\leq$ 0.4; blue symbols)
and perturbed ($HI-def$ $>$ 0.4; red symbols) galaxies. Empty symbols show the normalised distribution of equivalent diameters 
corrected for seeing effects whenever the correction is less than 50\%.
The dashed vertical line shows the equivalent diameter corresponding
to the mean seeing of the survey at the distance of the cluster (16.5 Mpc). The blue and red solid lines show the best fit 
to the normalised distributions for the unperturbed and perturbed samples measured in the range $100 \leq D_{eq} \leq 500$ pc.}
\label{Deffdist_all_lin}%
\end{figure}

Figure \ref{Deffdist_all_lin} shows the normalised distribution of the equivalent diameter $D_{eff}$ of \hii\ regions separately 
for perturbed and unperturbed systems. The two distributions are very similar: they can be both fitted with a 
$\log N \propto 0.0060 (\pm 0.0001)\times D_{eq}$ relation when limited in the diameter range $100 \leq D_{eq} \leq 500$ pc, 
suggesting that \hii\ regions have comparable sizes whenever they are located in \hi\ gas-rich or gas-poor systems.
We recall that the size of \hii\ regions is tightly connected to their dynamical age since it is related 
to the time that the region takes to expand within the ISM (Ambrocio-Cruz et al. 2016). Figure \ref{Deffdist_all_lin} thus also suggests that 
the dynamical age distribution of \hii\ regions is comparable in both samples.

\subsubsection{Composite electron density distribution}

\begin{figure}
\centering
\includegraphics[width=0.49\textwidth]{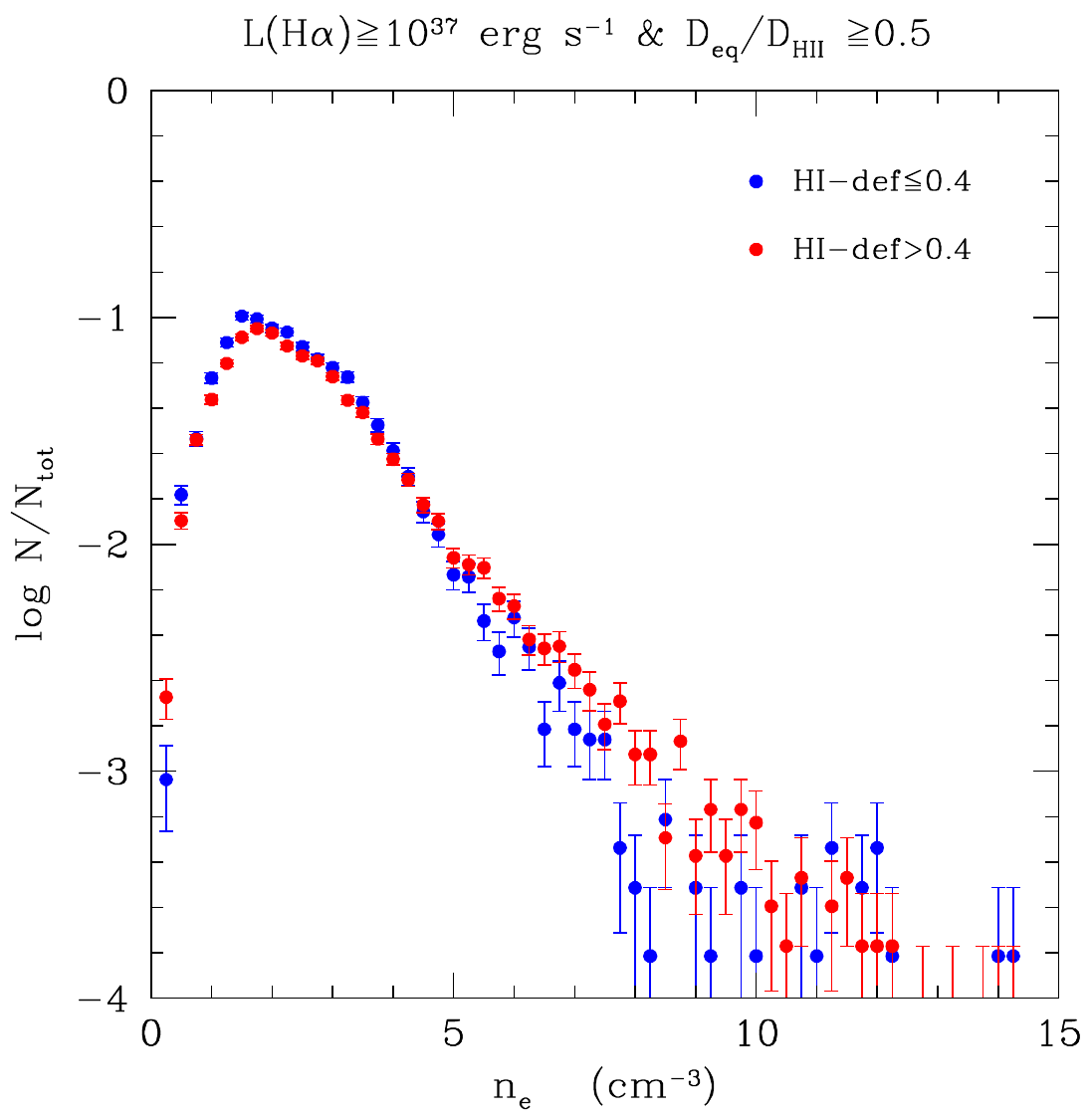}\\
\caption{Normalised distribution of the electron density derived using equivalent diameters corrected for seeing effects 
for {\hii} regions with $L(H\alpha)$ $\geq$ 10$^{37}$ erg~s$^{-1}$ and a diameter correction factor $\leq$ 50\%
for unperturbed ($HI-def$ $\leq$ 0.4; blue symbols) and perturbed ($HI-def$ $>$ 0.4; red symbols) galaxies. }
\label{nedist_all}%
\end{figure}

\begin{figure}
\centering
\includegraphics[width=0.49\textwidth]{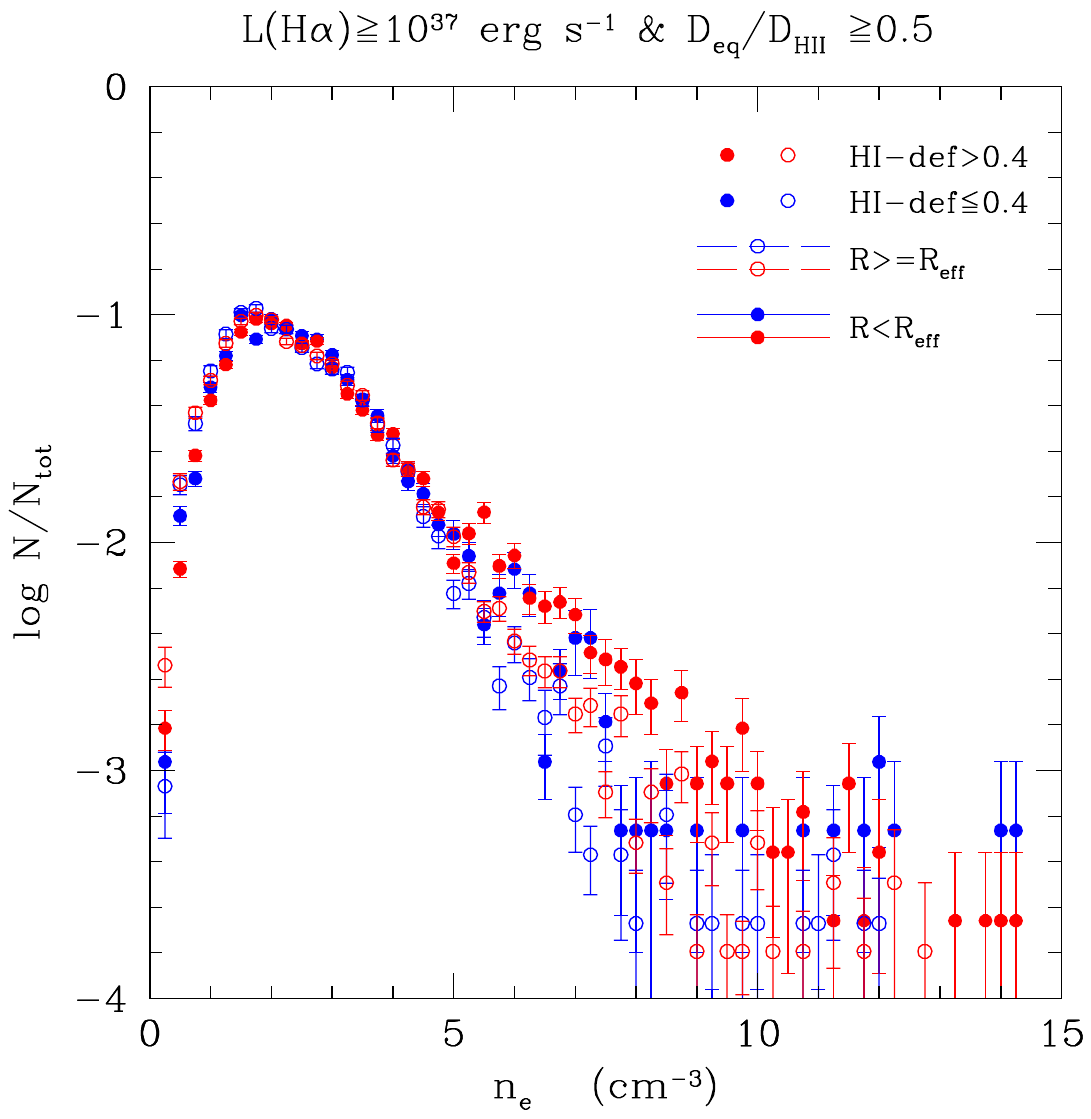}\\
\caption{Normalised distribution of the electron density derived using equivalent diameters corrected for seeing effects 
for {\hii} regions with $L(H\alpha)$ $\geq$ 10$^{37}$ erg~s$^{-1}$ and a diameter correction factor $\leq$ 50\%\ detected within (filled dots, solid lines)
and outside (empty dots, dashed lines) the $i$-band effective radius
for unperturbed ($HI-def$ $\leq$ 0.4; blue symbols) and perturbed ($HI-def$ $>$ 0.4; red symbols) galaxies. }
\label{nedistReff_all}%
\end{figure}

Figure \ref{nedist_all} shows the normalised distribution of the electron density defined using eq. 2 and equivalent diameters corrected for seeing effects 
for {\hii} regions with $L(H\alpha)$ $\geq$ 10$^{37}$ erg~s$^{-1}$ and a diameter correction factor $\leq$ 50\% for 
unperturbed and perturbed systems. Contrary to what is observed for the equivalent diameter, the two distributions are significantly different 
(a Kolmogorov-Smirnov test gives a probability $p$-value = 5$\times$10$^{-5}$ that they are drawn from the same distribution). 
Gas-poor, perturbed systems have a larger fraction of dense ($n_e$ $\gtrsim$ 5~cm$^{-3}$) and a lower fraction of low density 
($n_e$ $\lesssim$ 5~cm$^{-3}$) \hii\ regions than unperturbed systems: the mean of the ratio of the electron density distributions of unperturbed to perturbed systems is 
1.08$\pm$0.19 for $n_e$ $<$ 5 cm$^{-3}$ and 0.65$\pm$0.62 for $n_e$ $\geq$ 5 cm$^{-3}$. The densest regions with $n_e$ $\gtrsim$ 5 cm$^{-3}$ are mainly 
located within the inner regions, in particular in perturbed systems (Fig. \ref{nedistReff_all}).

\subsubsection{Luminosity - diameter relation}

\begin{figure}
\centering
\includegraphics[width=0.49\textwidth]{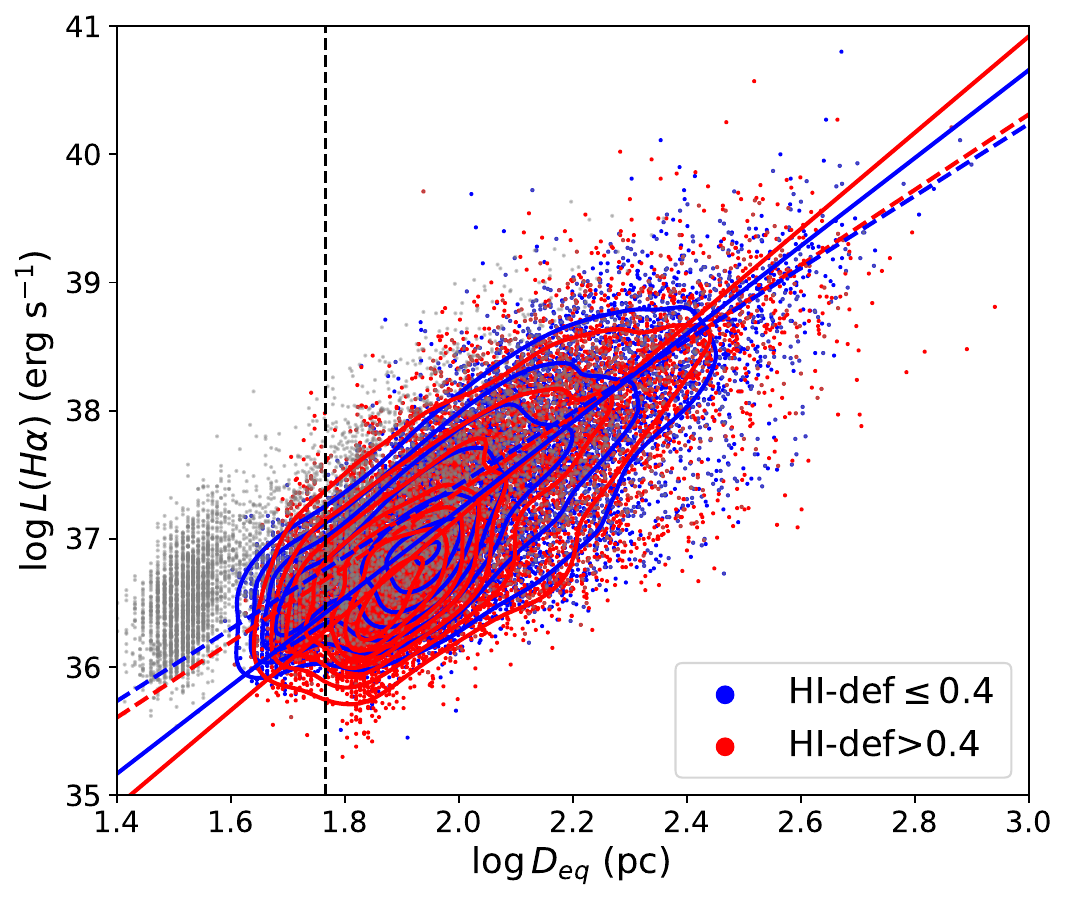}\\
\caption{Relation between the H$\alpha$ luminosity of individual {\hii} regions corrected for [{\nii}] contamination and dust 
attenuation and the equivalent diameter corrected for seeing effects. Blue and red filled dots are for {\hii} regions  
with a diameter correction factor $\leq$ 50\% for unperturbed ($HI-def$ $\leq$ 0.4; blue symbols) and perturbed ($HI-def$ $>$ 0.4; red symbols) galaxies,
respectively. Grey dots are for all {\hii} regions with no limits in diameter correction. 
The vertical dashed line shows the mean FWHM of the survey
assuming galaxies at the distance of the main body of the cluster (16.5 Mpc). The blue and red solid lines give the 
bisector fit (Isobe et al. 1990; see Table \ref{Tablumsize}) of the relation
derived for {\hii} regions with a diameter correction factor $\leq$ 50\%\ (blue and red dots), the dashed lines
for {\hii} regions with no limits in the diameter correction.
}
\label{sizelumall}%
\end{figure}

\begin{table}
\caption{Best-fit parameters for the luminosity-size relation (log $L(H\alpha)$ = $a$ $\times$ log $D_{eq}$ + $b$).
}
\label{Tablumsize}
{\scriptsize
\[
\begin{tabular}{cccccc}
\hline
\noalign{\smallskip}
\hline
Sample					& N.objects		&  $a$  	& $b$   	& $\rho$	& $\sigma$ 	\\
\hline
Unperturbed				&			&		&		&	&		\\
\hline
All$^a$					& 18\,474		& 2.82$\pm$0.01	& 31.79$\pm$0.04& 0.79	& 0.17   	 \\		 
$D_{eq}/D_{HII}$ $\geq$ 50\%		& 11\,263		& 3.43$\pm$0.02	& 30.36$\pm$0.05& 0.76	& 0.14   	 \\
\hline
Perturbed				&			&		&		&	&		\\
\hline
All$^a$					& 32\,379		& 2.94$\pm$0.01	& 31.49$\pm$0.03& 0.66	& 0.18   	 \\		 
$D_{eq}/D_{HII}$ $\geq$ 50\% 		& 21\,672		& 3.76$\pm$0.01	& 29.65$\pm$0.04& 0.69	& 0.14   	 \\
\hline
\end{tabular}
\]
Notes: Fits are done for all {\hii} regions with a diameter correction factor $\leq$ 50\%, unless otherwise stated. No limits on the H$\alpha$
luminosity is taken.
$\rho$ is the Spearman correlation coefficient, $\sigma$ the dispersion perpendicular to the fitted relation.\\
$a$ : no limits in the diameter correction.
}
\end{table}

Figure \ref{sizelumall} shows the luminosity vs. size relation of \hii\ regions located within perturbed and unperturbed galaxies.
The two variables are strongly correlated in both samples (see Table \ref{Tablumsize}), 
although with relations which appear statistically different in slope and intercept, with a steeper relation observed in \hi\ gas-deficient galaxies 
vs. \hi\ gas-rich systems.

\subsection{Scaling relations}

As in Boselli et al. (2025), we derive the main scaling relations characterising the statistical properties of \hii\ regions for 
unperturbed and perturbed systems and compare them to identify and quantify any possible effect of the different kind of interactions on their
physical and statistical properties. For this exercise we limit the comparison to those relations
identified in Boselli et al. (2025) as the most representative to trace the scaling properties of individual \hii\ regions in unperturbed systems.
We also focus the analysis to the differences between unperturbed and perturbed systems, and refer the reader interested to the
physical interpretation of these relations to Boselli et al. (2025).

\begin{figure}
\centering
\includegraphics[width=0.49\textwidth]{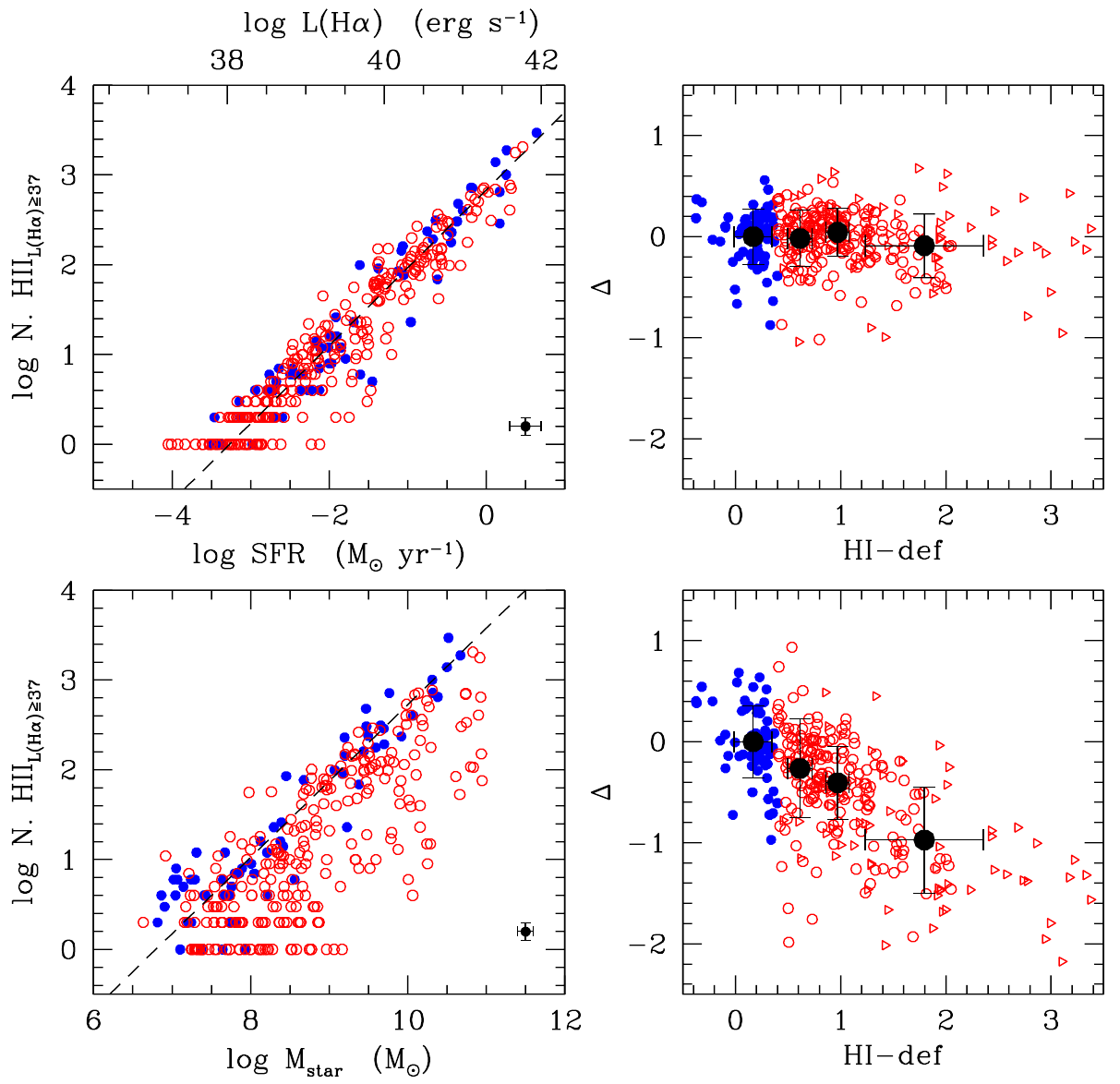}\\
\caption{Relation between the total number of {\hii} regions detected by \textsc{HIIphot} with $L(H\alpha)$ $\geq$ 10$^{37}$ erg~s$^{-1}$ corrected 
for dust attenuation and [{\nii}] contamination and the total star
formation rate (upper left panel) and total stellar mass (lower left) of the host galaxies. Star formation rates (lower axis) have been derived from 
H$\alpha$ luminosities (upper axis) corrected for dust attenuation and [{\nii}] contamination assuming a Chabrier (2003) IMF.
Blue filled dots are for unperturbed systems with a normal \hi\ gas content ($HI-def$ $\leq$ 0.4), red empty circles
gas-deficient perturbed galaxies ($HI-def$ $>$ 0.4). The black dot in the lower right corner shows the typical error bar in the data.
The black dashed line shows the best fit to the data (bisector fit) derived for galaxies with $HI-def$ $\leq$ 0.4 
(the best fit parameters are given in Table 5 of paper XVII).
The right panels show the relation between the dispersion of the scaling relations and the \hi\--deficiency parameter. 
Open triangles are for \hi\--undetected galaxies (lower limits to the \hi\--deficiency parameter). The large black filled dots with their error-bars 
show mean values and their dispersion in different bins of \hi\--deficiency. 
}
\label{scaling1N37all}%
\end{figure}

\subsubsection{Scaling relations characterising the statistical properties of \hii\ regions}

Figure \ref{scaling1N37all} shows the relation between the total number of {\hii} regions with $L(H\alpha)$ $\geq$ 10$^{37}$ erg~s$^{-1}$ and the total star
formation rate ($SFR$; upper panels) and total stellar mass ($M_{star}$; lower panels) of the host galaxies. The upper left panel of Fig. \ref{scaling1N37all} clearly
shows that unperturbed \hi\ gas-rich and perturbed gas-poor systems share the same relation, suggesting that the contribution of individual \hii\ regions
to the total star formation rate of galaxies is similar in the two populations. 
On the contrary, perturbed systems host, on average, a significantly 
smaller number of \hii\ regions (a factor of $\simeq$ 5-10) at a given stellar mass of the host galaxy than their gas-rich counterparts. 
It is also clear that the number of \hii\ regions per unit 
stellar mass strongly decreases with the \hi\--deficiency parameter, i.e. that the number of star forming regions in a galaxy is strongly related 
to its total mass of the cold gas reservoir.
These results indicate that the difference in the mean number of {\hii} regions with $L(H\alpha)$ $\geq$ 10$^{37}$ erg~s$^{-1}$ observed 
in the perturbed and unperturbed samples (see Sec. 3) is not due to a selection bias but it is a real physical effect related to a 
reduced star formation activity in \hi-deficient objects.  

\begin{figure}
\centering
\includegraphics[width=0.49\textwidth]{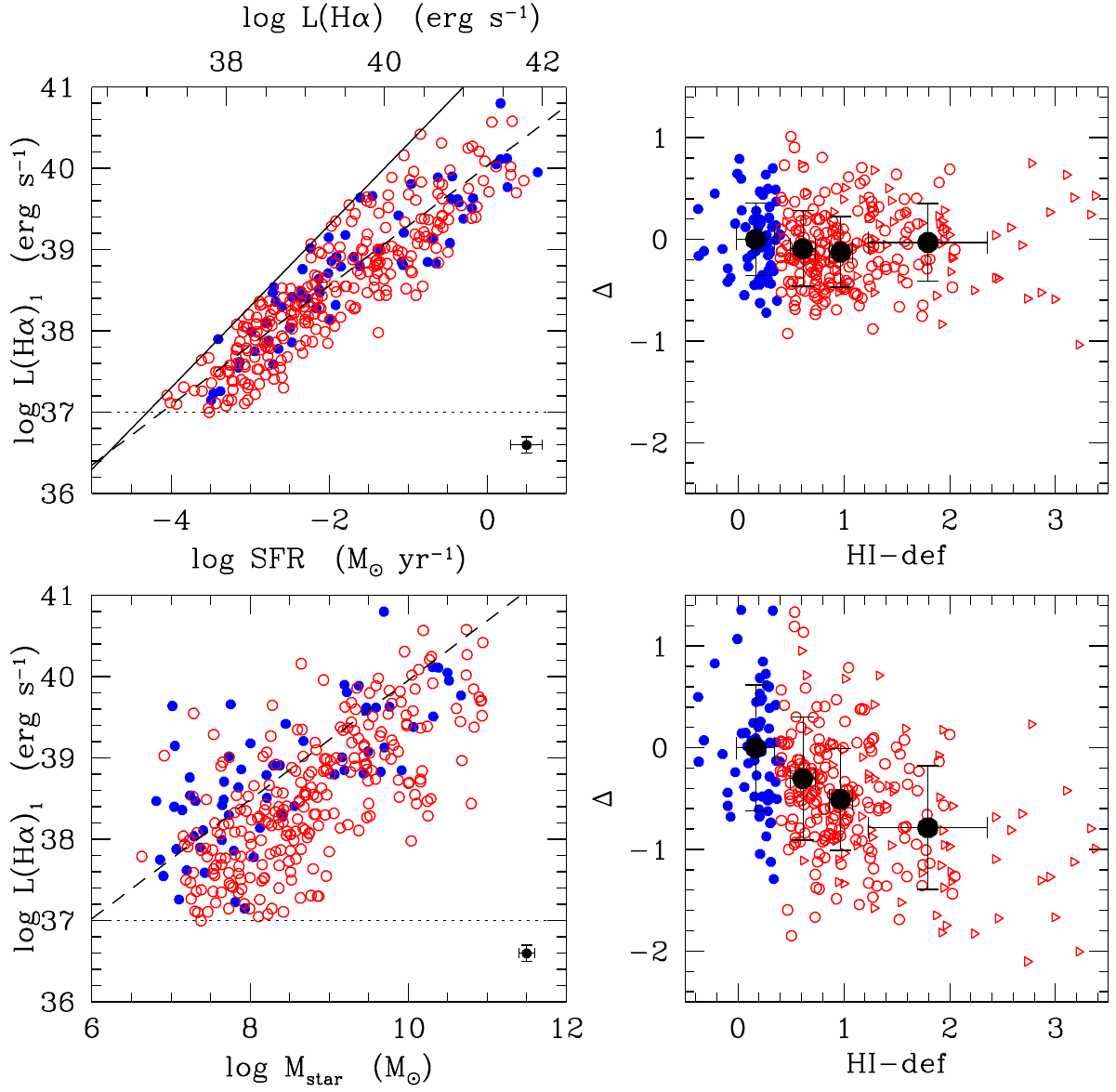}\\
\caption{Relation between the H$\alpha$ luminosity of the brightest {\hii} region corrected 
for dust attenuation and [{\nii}] contamination and the total star
formation rate (upper left panel) and total stellar mass (lower left) of the host galaxies. Star formation rates (lower axis) have been derived from 
H$\alpha$ luminosities (upper axis) corrected for dust attenuation and [{\nii}] contamination assuming a Chabrier (2003) IMF.
Blue filled dots are for unperturbed systems with a normal \hi\ gas content ($HI-def$ $\leq$ 0.4), red empty circles
gas-deficient perturbed galaxies ($HI-def$ $>$ 0.4), red empty triangles lower limits to the \hi\--deficiency parameter. 
The black dot in the lower right corner shows the typical error bar in the data.
The black dashed line shows the best fit to the data (bisector fit) derived for galaxies with $HI-def$ $\leq$ 0.4 
(the best fit parameters are given in Table 5 of paper XVII).
The right panels show the relation between the dispersion of the scaling relations and the \hi\--deficiency parameter. 
The large black filled dots with their error-bars show mean values and their dispersion in different bins of \hi\--deficiency. 
}
\label{scaling1Lupall}%
\end{figure}

Figure \ref{scaling1Lupall} shows the relation between the H$\alpha$ luminosity of the first ranked \hii\ region and the star formation rate 
(upper panels) and stellar mass (lower panels) of the host galaxies. As for Fig. \ref{scaling1N37all}, we do not see any significant difference
between perturbed and unperturbed systems in the relation with the star formation activity, while a clear segregation is evident in the 
relation with stellar mass, where again the typical H$\alpha$ luminosity of the brightest \hii\ region in galaxies of comparable stellar mass 
is higher in unperturbed systems than in perturbed objects. As above, the offset from this relation is strongly correlated with the 
\hi\--deficiency parameter, suggesting again that the properties of the brightest \hii\ region depends on the total atomic gas reservoir of
the host galaxy. It is worth noticing here that the same behaviour is observed whenever the mean H$\alpha$ luminosity of the first three ranked
\hii\ regions is considered (see Fig. \ref{scaling1Lup3all}).

\begin{figure}
\centering
\includegraphics[width=0.49\textwidth]{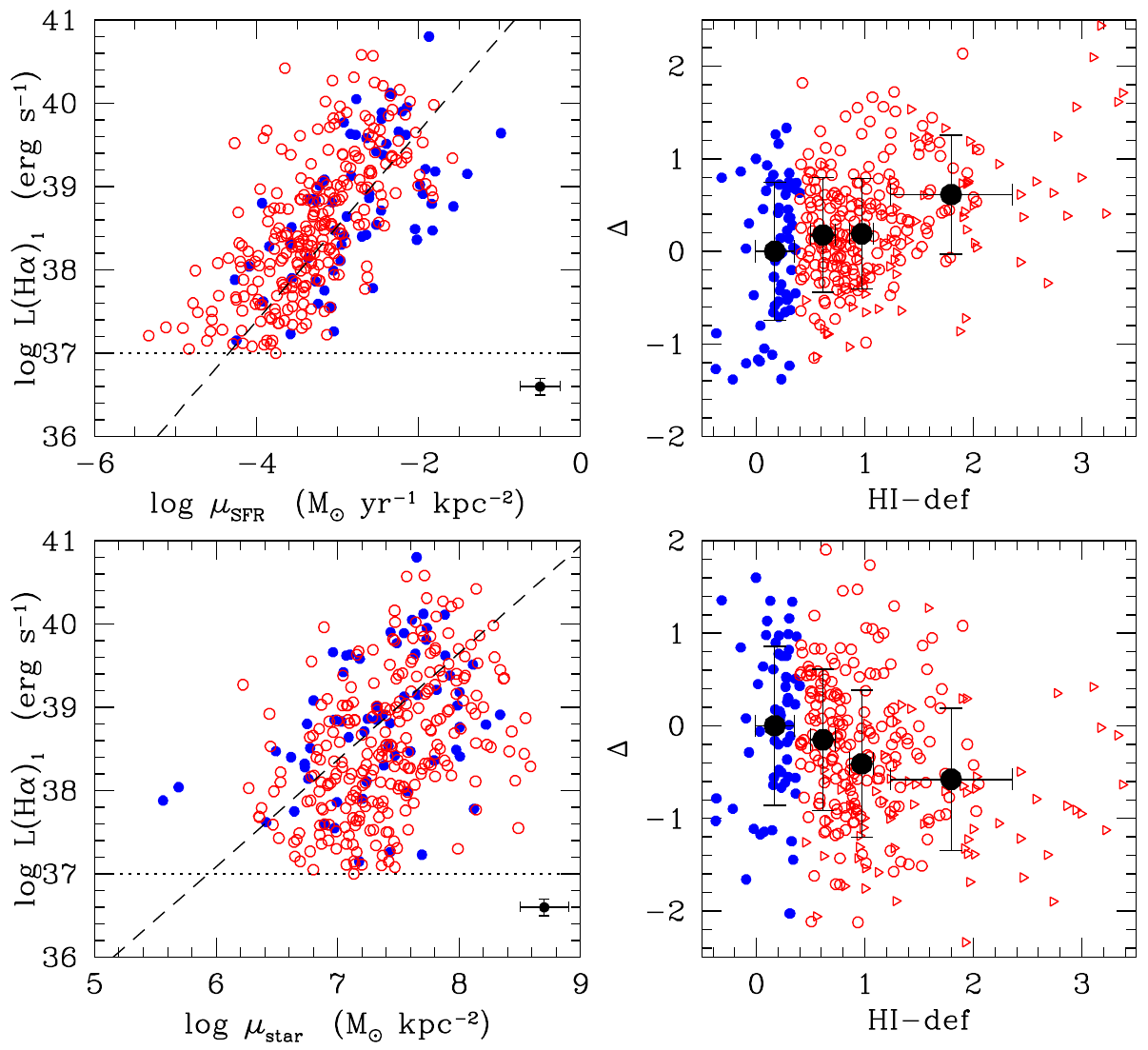}\\
\caption{Relation between the H$\alpha$ luminosity of the brightest {\hii} region corrected 
for dust attenuation and [{\nii}] contamination and the total star
formation rate (upper left panel) and total stellar mass (lower left) surface densities of the host galaxies. 
Blue filled dots are for unperturbed systems with a normal \hi\ gas content ($HI-def$ $\leq$ 0.4), red empty circles
gas-deficient perturbed galaxies ($HI-def$ $>$ 0.4), red empty triangles lower limits to the \hi\--deficiency parameter. 
The black dot in the lower right corner shows the typical error bar in the data.
The black dashed line shows the best fit to the data (bisector fit) derived for galaxies with $HI-def$ $\leq$ 0.4 
(the best fit parameters are given in Table 5 of paper XVII).
The right panels show the relation between the dispersion of the scaling relations and the \hi\--deficiency parameter. 
The large black filled dots with their error-bars show mean values and their dispersion in different bins of \hi\--deficiency. 
}
\label{scaling1LupSigmaall}%
\end{figure}

Figure \ref{scaling1LupSigmaall} shows the relationship between the H$\alpha$ luminosity of the brightest \hii\ region and the star formation surface density
and stellar mass surface density defined as in paper XVII. We recall that gas-rich and gas-poor galaxies have a similar size distribution (Fig. \ref{mstardist}).
Perturbed and unperturbed systems share the same relation, where the H$\alpha$ luminosity
of the brightest \hii\ region increases with increasing stellar mass and star formation rate surface density. We see
a clear trend between the dispersion of the relation and the \hi\--deficiency parameter, with \hi\--deficient galaxies having, on average,
brighter H$\alpha$ luminosities per unit star formation rate surface density and weaker luminosities per stellar mass surface density than their unperturbed counterparts.

\begin{figure}
\centering
\includegraphics[width=0.45\textwidth]{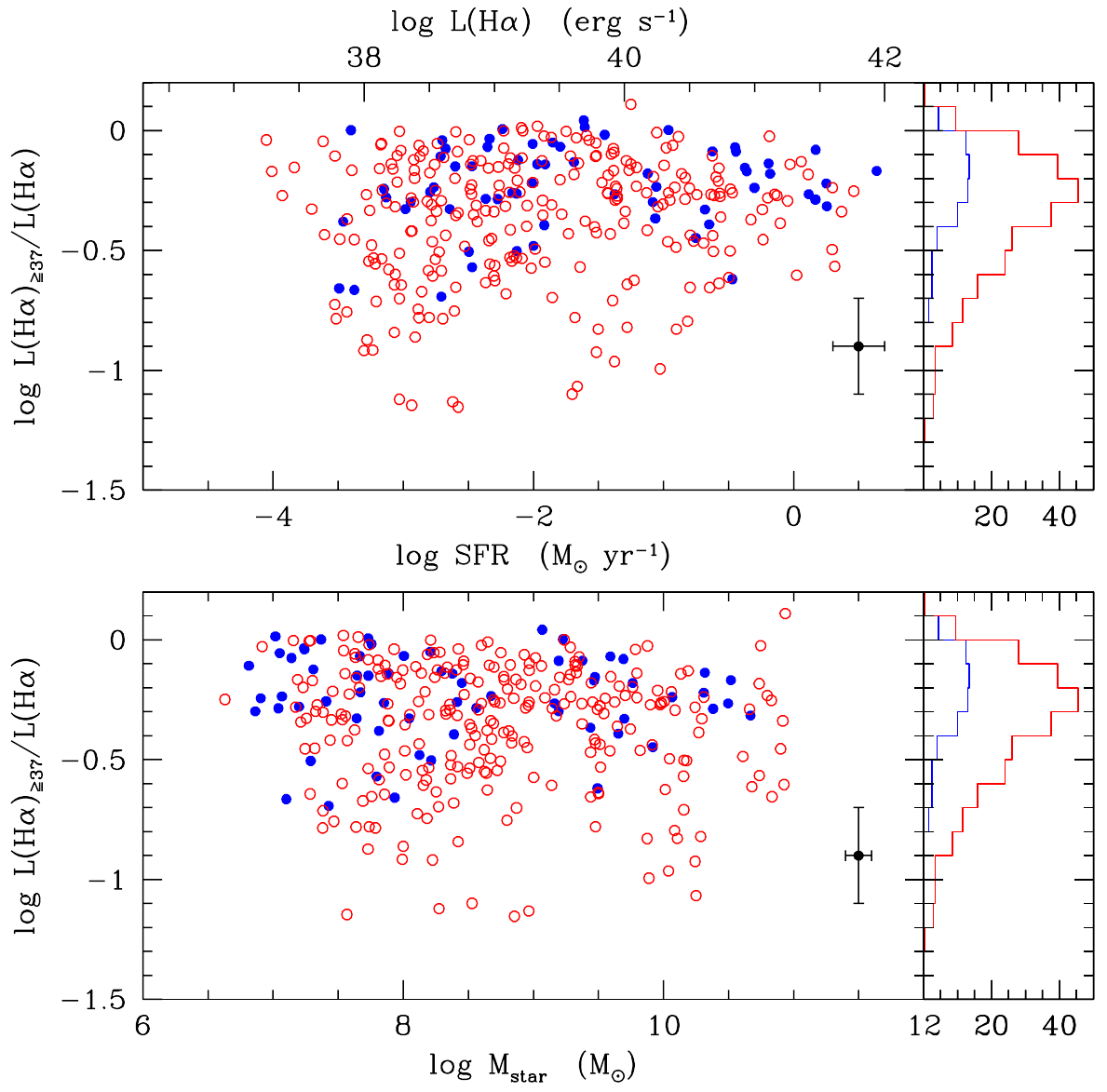}
\includegraphics[width=0.45\textwidth]{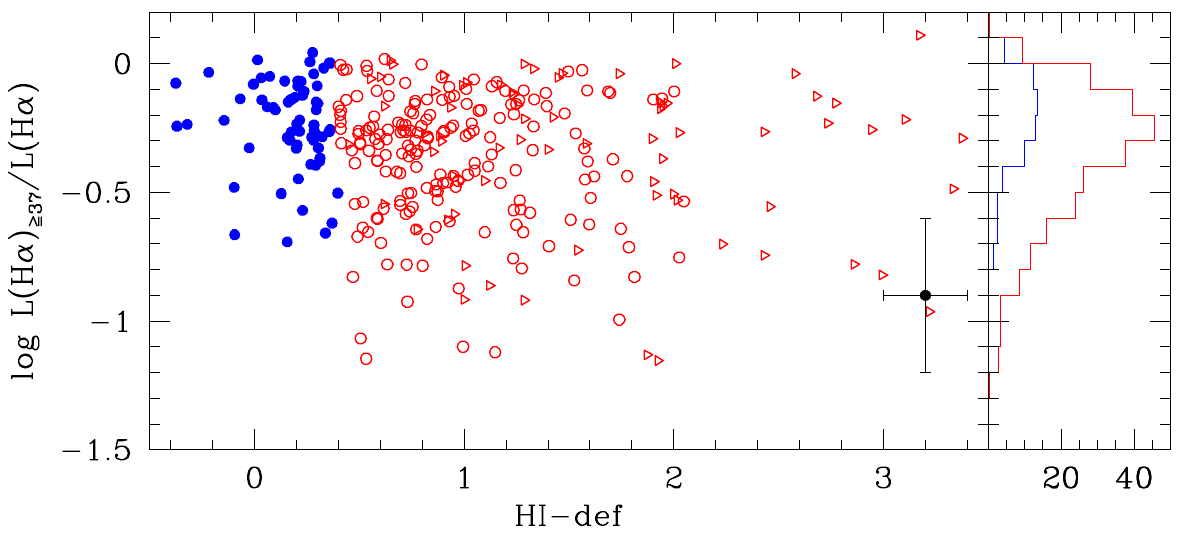}\\
\caption{Relation between $L(H\alpha)_{\geq 37}/L(H\alpha)$, defined as the ratio between the sum of the H$\alpha$ luminosity of all the {\hii} regions 
with luminosity $L(H\alpha)$ $\geq$ 10$^{37}$ erg~s$^{-1}$ and the integrated H$\alpha$ luminosity of the galaxy, and the total star formation rate  
(upper left panel), total stellar mass (middle left), and \hi\--deficiency parameter (lower left) of the host galaxies. Both quantities are corrected 
for dust attenuation and [{\nii}] contamination.
Blue filled dots are for unperturbed systems with a normal \hi\ gas content ($HI-def$ $\leq$ 0.4), red empty circles
gas-deficient perturbed galaxies ($HI-def$ $>$ 0.4), red empty triangles lower limits to the \hi\--deficiency parameter. 
The black dot in the lower right corner shows the typical error bar in the data.
The right panels show the distribution of the \hi\--normal and \hi\--deficient galaxy populations. 
}
\label{scaling1L37all}%
\end{figure}

Figure \ref{scaling1L37all} shows the relation between $L(H\alpha)_{\geq 37}/L(H\alpha)$ (ratio between the sum of the H$\alpha$ luminosity of all the {\hii} regions 
with luminosity $L(H\alpha)$ $\geq$ 10$^{37}$ erg~s$^{-1}$ and the integrated H$\alpha$ luminosity of the galaxy, this last including the diffuse emission
within the optical diameter of the galaxy) and the total star formation rate, 
stellar mass, and \hi\--deficiency parameter of the host galaxies. The different variables are not correlated. We observe, however, a skewed distribution
of $L(H\alpha)_{\geq 37}/L(H\alpha)$ which extends to lower values in perturbed vs. unperturbed systems. A Kolmogorov-Smirnov test indicates that 
the two distributions are statistically different ($p$-value = 2 $\times$ 10$^{-3}$). This suggests that the contribution of 
the diffuse H$\alpha$ emission is more important in perturbed, gas-deficient galaxies than in unperturbed systems.

\begin{figure}
\centering
\includegraphics[width=0.45\textwidth]{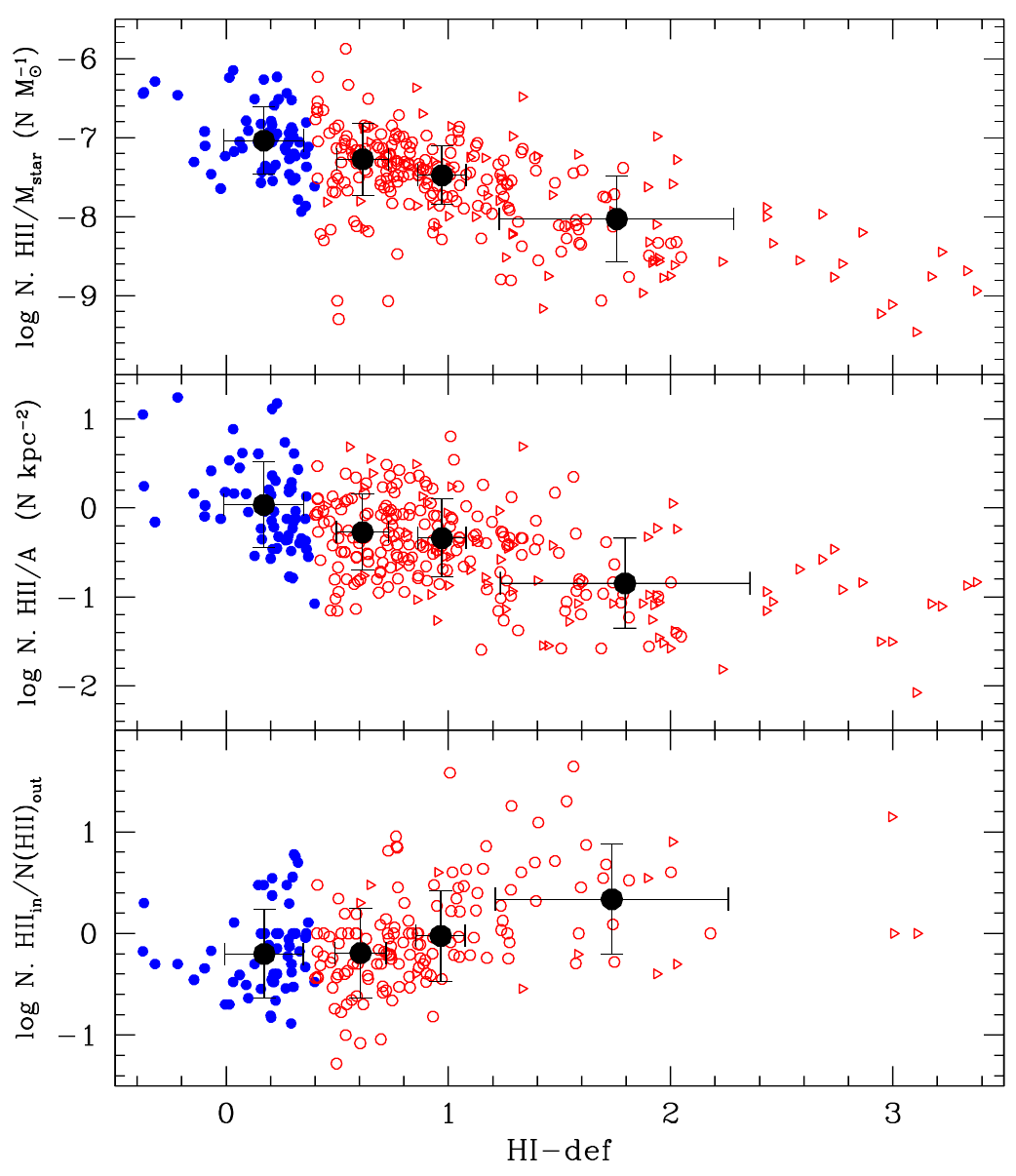}\\
\caption{Relation between the number of {\hii} regions of luminosity $L(H\alpha)$ $\geq$ 10$^{37}$ erg~s$^{-1}$ corrected 
for dust attenuation and [{\nii}] contamination per unit stellar mass disc, (upper panel) 
stellar disc surface (measured up to the 25.5 $B$-band isophotal diameter; central panel), and the ratio of total number of {\hii} regions located 
inside and outside the $i$-band effective radius (in log scale; lower panel) and the \hi\--deficiency parameter.
Blue filled dots are for unperturbed systems with a normal \hi\ gas content ($HI-def$ $\leq$ 0.4), red empty circles
gas-deficient perturbed galaxies ($HI-def$ $>$ 0.4). The large black filled dots with their error-bars show mean values and their dispersion in 
different bins of \hi\--deficiency. 
}
\label{ratiodensityHIdef}%
\end{figure}

Figure \ref{ratiodensityHIdef} shows the relationship between the number of \hii\ regions per unit stellar mass (upper panel), per unit disc surface
(central panel), and the ratio of \hii\ regions detected within and outside the $i$-band effective radius of the host galaxy (lower panel) with 
the \hi\--deficiency parameter. The three variables are clearly correlated with the total \hi\ gas content of the host galaxy and consistently suggest
that gas-poor systems have, on average, a lower number of \hii\ regions per unit stellar mass or stellar disc area than gas-rich galaxies, and that
they lack of \hii\ regions mainly in the outer disc (see Boselli et al. 2020). As for Fig. \ref{scaling1N37all}, this is a further evidence 
that the observed decrease of the number of \hii\ regions in perturbed galaxies is physical and not due to selection effects.

\subsubsection{Fit parameters of the luminosity function on individual galaxies}

\begin{figure}
\centering
\includegraphics[width=0.45\textwidth]{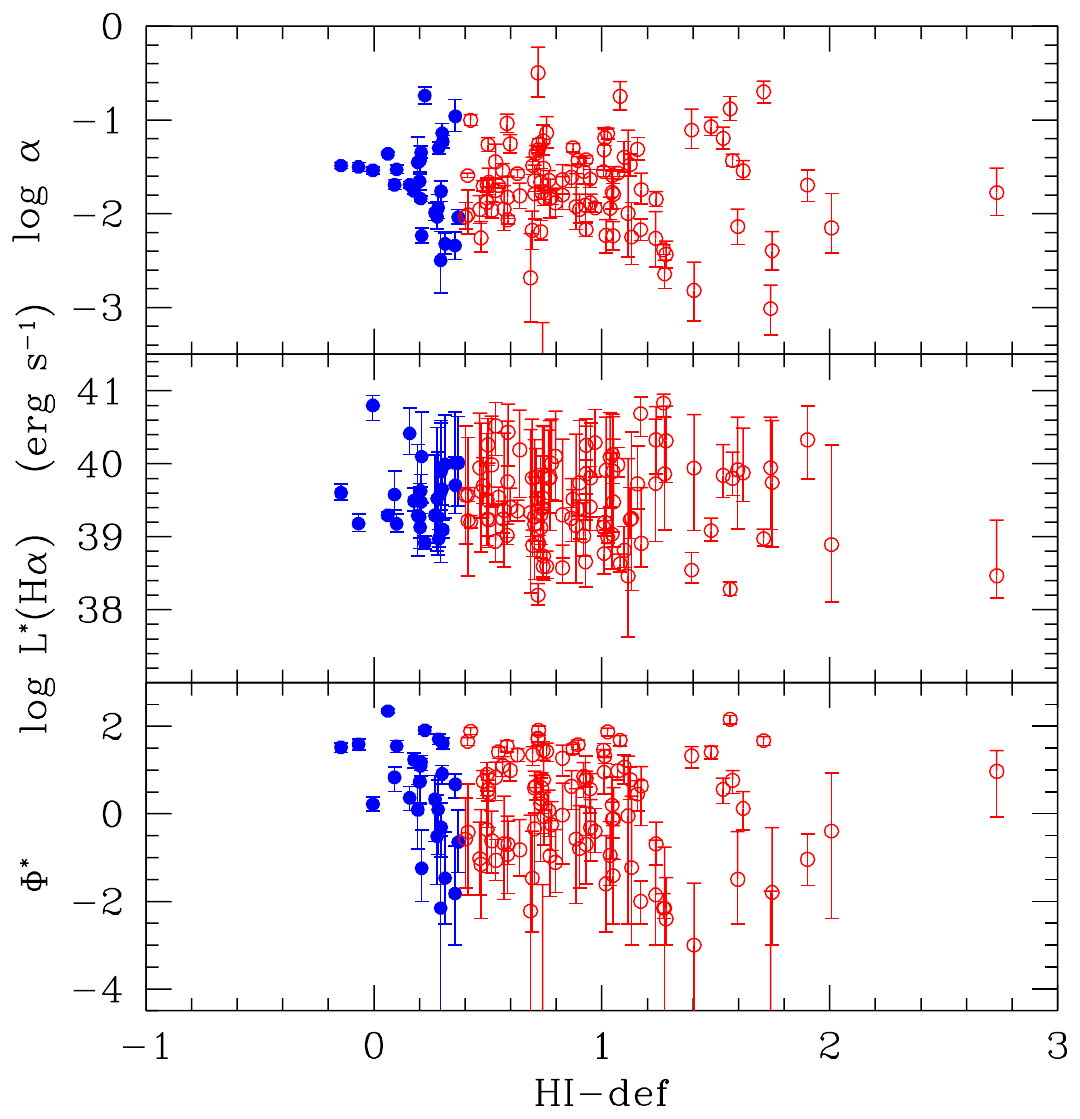}\\
\caption{Relation between the output parameters ($\alpha$, upper panel; $L^*(H\alpha)$, central panel; $\Phi^*$, lower panel) of the best fit Schechter 
function on individual galaxies with more than 20 \hii\ regions of H$\alpha$ luminosity brighter than $L(H\alpha)$ $\geq$ 10$^{37}$ erg~s$^{-1}$ and the \hi\--deficiency parameter.
Blue filled dots are for unperturbed systems with a normal \hi\ gas content ($HI-def$ $\leq$ 0.4), red empty circles
gas-deficient perturbed galaxies ($HI-def$ $>$ 0.4). 
}
\label{fitLFdef}%
\end{figure}

The sample analysed in this work includes 127 galaxies with more than 20 \hii\ regions of luminosity $L(H\alpha)$ $\geq$ 10$^{37}$ erg~s$^{-1}$ each once 
corrected for [{\nii}] contamination and dust attenuation. For these galaxies we can fit a Schechter function and derive the best fit parameters 
characterising their luminosity function, significantly increasing the statistics of the unperturbed galaxy sample which is limited to 27 objects. 
Figure \ref{fitLFdef} shows the relation between the output parameters of the best fit and the \hi\--deficiency parameter. There is no clear relation 
between the different variables, suggesting that the lack of atomic gas and/or the importance of the undergoing perturbation does not affect the 
H$\alpha$ luminosity function of \hii\ regions in individual galaxies. This result suggests that the only parameter of the 
luminosity function affected by external perturbations is their total number. At a first glance, this 
result looks in contradiction with what shown in Fig. \ref{LFHIIall}, i.e. that the faint end slope and the characteristic luminosity
of the two galaxy populations are significantly different (see Sec. 4.1.1). We notice, however, that
the variations observed in the best fit parameters of the composite luminosity function of perturbed and unperturbed systems (see Table \ref{TabLF}) 
are very small when compared with the wide range of parameters sampled by the best fit of individual galaxies.

\section{Discussion and conclusion}

The analysis presented in the previous section shows that perturbed and unperturbed galaxies are characterised by \hii\ regions with statistical and 
physical properties slightly different. Their composite H$\alpha$ luminosity function can be fairly well represented by a Schechter function, but with 
characteristic parameters statistically significantly different.
In perturbed systems, the composite \hii\ luminosity function has a steeper faint-end slope and a brighter characteristic luminosity than in unperturbed 
objects (Fig. \ref{LFHIIall}). The difference in the two distributions comes principally from the outer regions, those located outside the 
effective radius (Fig. \ref{LFHIIReffallnew} and \ref{ratiodensityHIdef}). We do not observe, 
however, any significant difference in the best fit parameters of individual galaxies (Table \ref{TabLF}; Fig. \ref{fitLFdef}). 
The composite \hii\ size distribution is similar in the two samples (Fig. \ref{Deffdist_all_lin}), 
but the H$\alpha$ size-luminosity relation is steeper in perturbed systems vs. unperturbed objects (Fig. \ref{sizelumall}). For this reason, the two galaxy populations have also 
different composite mean electron density distributions, with \hi\--deficient systems hosting a larger fraction of high-density regions than gas-rich galaxies
(Fig. \ref{nedist_all}). More specifically, the densest \hii\ regions ($n_e$ $\gtrsim$ 5 cm$^{-3}$) are mainly located in the inner disc (Fig. \ref{nedistReff_all}) where the stellar 
density is higher. This is expected since the star formation process is triggered by the midplane pressure produced
by the stellar gravity of the disc (Shi et al. 2018). The increased pressure favors the transformation of atomic to molecular gas and the formation of giant molecular clouds (Blitz
\& Rosolowsky 2006). Interesting is the fact that these densest regions are principally located in perturbed systems, possibly because these
objects are also suffering an increase of pressure due to their interaction with the surrounding environment. This result also
suggests that in perturbed objects the low-density ionised gas
is removed with the other gas phases during the interaction with the surrounding environment.

Overall, the \hii\ regions located within these two galaxy populations follow similar scaling relations but with several statistically significant 
differences: perturbed systems have, on average, a lower number of \hii\ regions per unit stellar mass (Fig. \ref{scaling1N37all}), 
or galaxy surface (Fig. \ref{ratiodensityHIdef}) than unperturbed objects, 
and the difference between the two galaxy populations increases with their \hi\--deficiency parameter. This is also the case for the H$\alpha$ 
luminosity of the first ranked (or first three ranked) \hii\ regions, which is, on average, lower in perturbed systems than in unperturbed galaxies 
of similar stellar mass (Fig. \ref{scaling1Lupall}, \ref{scaling1Lup3all}) and stellar mass surface density (Fig. \ref{scaling1LupSigmaall}), 
with differences which increase with the \hi\--deficiency parameter. On the contrary, the 
same variables (total number of \hii\ regions, H$\alpha$ luminosity of the first ranked and first three ranked \hii\ regions) at a given star formation 
rate and star formation rate surface density are similar in the two samples. Interesting is also the fact that the contribution of \hii\ regions to 
the total H$\alpha$ emission of galaxies is more important in unperturbed vs. perturbed systems, these last often characterised 
by a relevant diffuse ionised gas emission (Fig. \ref{scaling1L37all}). 

The observed differences in the radial distribution of \hii\ regions can be easily explained in a RPS scenario, where the gas is 
removed outside-in during the interaction of gas-rich, fresh-infalling galaxies with the hot and diffuse ICM (Gunn \& Gott 1972; Boselli et al. 2022b). 
The lack of gas, which is principally removed in the outer galaxy discs, prevents the formation of new stars, producing truncated discs in the young 
stellar populations (e.g. Boselli et al. 2006; see Fig. \ref{Haimage}). This truncation has been reported in different tracers such as ionised gas (Koopmann \& Kenney 2004a,b, 2006; 
Boselli \& Gavazzi 2006; Fossati et al. 2013; Boselli et al. 2015; Morgan et al. 2024), cold atomic (Warmels 1986, Cayatte et al. 1990, 1994; 
Bravo-Alfaro et al. 2000; Vollmer et al. 2001; Chung et al. 2009; Loni et al. 2021), and dust (Cortese et al. 2010, 2014). Recent CO observations
have consistently shown that the molecular hydrogen disc, the gas component which directly fuels star formation, of \hi-deficient 
galaxies is also truncated, and that 
their molecular gas content is, on average, lower than that of gas-rich systems (Fumagalli et al. 2009; Boselli et al. 2014b; 
Mok et al. 2017; Villanueva et al. 2022; Zabel et al. 2022). Interestingly, some of these results have been obtained using the CO data 
gathered during the VERTICO survey of the Virgo cluster (Brown et al. 2021, 2023; Jimenez-Donaire et al. 2023), thus using the same galaxies analysed in this work.
The size of the disc is reduced proportionally to the 
quantity of removed gas (e.g. Vollmer et al. 2001), explaining the strong dependence on the \hi\--deficiency parameter described above (see also Boselli 
et al. 2020). The increase of the faint-end slope of the H$\alpha$ luminosity function observed in perturbed systems might be reflecting the fact that in 
these galaxies the activity of star formation is gradually turning off because locally the gas content is reduced, producing an increasing number of low luminosity \hii\ regions.
It is worth mentioning that a steepening of the luminosity function in the outer disc has been also observed in the PHANGS sample of Santoro et al. (2022), as well as in the 
molecular clouds mass function (Braine \& Corbelli 2026).
The overall differences between perturbed and unperturbed systems, however, are moderate, thus these results suggest that the way stars (and thus the IMF) 
are formed in \hii\ regions is not largely affected by the large scale gas distribution. If analysed in more detail, however, this is only partly true
since we observe a mild increase of the ionised gas density in the \hii\ regions located in the inner discs of perturbed systems possibly resulting from 
the compression of the gas induced by their interaction with the surrounding ICM.

The increasing fraction of dense \hii\ regions observed in perturbed galaxies should be confirmed with independent and more direct tracers such as 
spectroscopic ionised gas line ratio diagnostics (e.g. [{\sii}]$\lambda$6716/[{\sii}]$\lambda$6731, Osterbrock \& Ferland 2006), or tracers of other 
gas phases (atomic, molecular). This increase, if confirmed, is not unexpected since it has been already observed in perturbed systems or predicted by tuned simulations, 
where the external pressure is compressing the gas on the stellar disc locally inducing an increase of the star formation activity (e.g. IC 3476, 
Boselli et al. 2021; see also Fujita \& Nagashima 1999; Bekki \& Couch 2003; Henderson \& Bekki 2016; Nehlig et al. 2016; Steyrleithner et al. 2020; 
Troncoso-Iribarren et al. 2020; Liz\'ee et al. 2021; Zhu et al. 2024). 

Finally, of great interest is also the fact that the relative contribution of the diffuse ionised gas emission to the total H$\alpha$ luminosity 
increases in gas-deficient, perturbed galaxies. This result seems to confirm previous claims gathered thanks to the analysis of limited samples of 
cluster galaxies with integral field units (IFU) spectroscopic data (e.g. Fossati et al. 2019, Pedrini et al. 2022) and recently highlighted during 
the MAUVE project (Brown et al. 2026), which consistently suggest that after the radial truncation of the star formation activity of perturbed 
galaxies the ionised gas emission in the outer disc is dominated by evolved stars. 

In a broader context of galaxy evolution in a rich environment, these results are consistent with a ram pressure stripping scenario,
the one generally proposed to explain the quenched nature of star-forming galaxies in massive, nearby clusters (e.g. Boselli et al. 2022b).  In this scenario, 
the gas is stripped outside-in, quenching the activity of star formation in the outer disc of massive spirals while completely suppressing it in dwarf 
systems, where a residual star formation can still be present only in the nucleus (Boselli et al. 2008a, 2008b, Boselli \& Gavazzi 2014). The quenching process
occurs on relatively short timescales ($\lesssim$ 500 Myr, e.g. Boselli et al. 2022b), producing inverted colour gradients in massive spirals 
(Boselli et al. 2006) and transforming rotationally-supported, star-forming low-mass discs into quiescent dwarf ellipticals (e.g. Boselli \& Gavazzi 2014).

\section{Data availability}

The full version of the Tables and of Fig. \ref{LFindividual} are available at \protect\url{https://mission.lam.fr/vestige/data.htm}. Tables
\ref{gal}, \ref{Tabds9}, \ref{galLF} are also available in electronic form at the CDS via anonymous ftp to cdsarc.u-strasbg.fr (130.79.128.5) or via http://cdsweb.u-strasbg.fr/cgi-bin/qcat?J/A+A/.

\begin{acknowledgements}

Based on observations obtained with
MegaPrime/MegaCam, a joint project of CFHT and CEA/DAPNIA, at the Canada-French-Hawaii Telescope
(CFHT) which is operated by the National Research Council (NRC) of Canada, the Institut National
des Sciences de l'Univers of the Centre National de la Recherche Scientifique (CNRS) of France and
the University of Hawaii. We are grateful to the whole CFHT team who assisted us in the preparation and in the execution of the observations and in the calibration and data reduction: 
Todd Burdullis, Daniel Devost, Bill Mahoney, Nadine Manset, Andreea Petric, Simon Prunet, Kanoa Withington.
We thank the anonymous referee for constructive comments and suggestions which helped improving the quality of the manuscript.
We are grateful to the whole CFHT team who assisted us in the preparation and in the execution of the observations and in the calibration and data reduction.
This work was supported by the Programme National Cosmology et Galaxies (PNCG) of CNRS/INSU with INP and IN2P3, co-funded by CEA and CNES.
This research has made use of the NASA/IPAC Extragalactic Database (NED) 
which is operated by the Jet Propulsion Laboratory, California Institute of 
Technology, under contract with the National Aeronautics and Space Administration
and of the GOLDMine database (http://goldmine.mib.infn.it/) (Gavazzi et al. 2003).
MB acknowledges support by the ANID BASAL project FB210003. This work was supported by the French government through the France 2030 investment plan managed 
by the National Research Agency (ANR), as part of the Initiative of Excellence of Universit\'e C\^ote d'Azur under reference No. ANR-15-IDEX-01. 
MB also acknowledges support from the French National Research Agency (ANR), grant ANR-24-CE92-0044 (project STARCLUSTERS).

\end{acknowledgements}

\begin{appendix}
\onecolumn

\section{Properties of the perturbed sample galaxies}

The following Tables give the main parameters used in the present analysis for the $perturbed~sampe$ and can be compared to those 
already presented for the $unperturbed~sample$ in Boselli et al. (2025). Table \ref{gal} lists all galaxies of the sample, Table \ref{Tabds9}
gives the parameters of the elliptical apertures used to identify the \hii\ regions and to measure total fluxes. Table \ref{galLF} provides
the best-fit parameters for the luminosity function of individual objects. Finally, we provide only in electronic format the output of the \textsc{HIIphot} 
code for the full sample ($perturbed$ and $unperturbed$ galaxies). The full set of data will be made available on CDS or on a 
dedicated VESTIGE webpage\footnote{https://mission.lam.fr/vestige/index.html}. 

\begin{landscape}
\begin{table}[h!]
\caption{Galaxies analysed in this work (perturbed sample)
}
\label{gal}
{\tiny 
\[
\resizebox{\columnwidth}{!}{
\begin{tabular}{cccccccccccccccc}
\hline
\noalign{\smallskip}
\hline
Name		& NGC/IC  & RA(J2000)   & DEC(J2000)    & $E(B-V)$        &$2a_{B25.5}$ &$R_{eff,i}$&$PA(i)$    & $b/a$   &log $M_{star}$&SM     &dist        &$A(H\alpha)$&[{\nii}]/H$\alpha$&$HI-def$&N{\hii}$_{L(H\alpha)\geq 37}$\\
		&	  &  h m s      & $^o$ ' "      &                 & arcmin     & arcsec     & deg       &         & M$_{\odot}$  &       & Mpc	      & mag	   &	      &           &       \\
\hline
    VCC15                   & 3021       & 12:09:54.58 & +13:03:00.0 & 0.0379   &  1.2    &   16.41    &     31     &     0.7   &   9.0     &      M     &    32.0  &	0.0   &     0.26   &	  0.74  &    22      \\   
    VCC34                   & 3033       & 12:11:09.96 & +13:35:15.1 & 0.0309   &  1.45   &   12.91    &     177    &     0.49  &   8.63    &      LVC   &    16.5  &	0.0   &     0.12   &	  0.7	&    13      \\   
    VCC41                   & -          & 12:12:04.39 & +12:44:28.1 & 0.027    &  0.76   &   11.21    &     173    &     0.5   &   8.28    &      M     &    32.0  &	0.0   &     0.08   &	  1.15  &    1       \\   
    VCC47                   & 4165       & 12:12:11.81 & +13:14:47.5 & 0.0351   &  1.26   &   14.51    &     156    &     0.75  &   9.82    &      M     &    32.0  &	1.01  &     0.41   &	  0.92  &    122     \\   
    VCC48                   & 3036       & 12:12:15.08 & +12:29:16.8 & 0.0287   &  2.14   &   21.97    &     175    &     0.49  &   8.65    &      LVC   &    16.5  &	0.13  &     0.12   &	  0.59  &    23      \\   
    VCC67                   & 3044       & 12:12:48.46 & +13:58:34.4 & 0.0371   &  2.82   &   19.42    &     72     &     0.72  &   8.84    &      LVC   &    16.5  &	0.05  &     0.16   &	  0.77  &    24      \\
\hline
\end{tabular}}
\]
}
\end{table}
\end{landscape}

\begin{landscape}
\setcounter{table}{0}
\begin{table}
\caption{continued
}
\label{gal}
{\tiny 
\[
\resizebox{\columnwidth}{!}{\begin{tabular}{cccccccccccccccc}
\hline
\noalign{\smallskip}
\hline
Name		& NGC/IC  & RA(J2000)   & DEC(J2000)    & $E(B-V)$        &$2a_{B25.5}$ &$R_{eff,i}$&$PA(i)$    & $b/a$   &log $M_{star}$&SM     &dist        &$A(H\alpha)$&[{\nii}]/H$\alpha$&$HI-def$&N{\hii}$_{L(H\alpha)\geq 37}$\\
		&	  &  h m s      & $^o$ ' "      &                 & arcmin     & arcsec     & deg       &         & M$_{\odot}$  &       & Mpc	      & mag	   &	      &           &       \\
\hline
    VCC381                  & -          & 12:19:53.47 & +06:39:56.9 & 0.0209   &  1.07   &   20.03    &     57     &     0.9   &   7.57    &      B     &    15.8  &	0.0   &     0.04   &	  0.68  &    3       \\   
    VCC393                  & 4276       & 12:20:07.49 & +07:41:30.7 & 0.0293   &  2.0    &   20.46    &     94     &     0.53  &   9.24    &      B     &    15.8  &	1.92  &     0.51   &	  0.78  &    178     \\   
    VCC409                  & -          & 12:20:23.16 & +06:08:05.7 & 0.0196   &  0.29   &   12.27    &     73     &     0.78  &   8.15    &      W     &    32.0  &	0.0   &     0.07   &	  0.62  &    1       \\   
    VCC413                  & -          & 12:20:23.31 & +13:53:53.0 & 0.0392   &  0.54   &   5.13     &     121    &     0.43  &   7.47    &      LVC   &    16.5  &	0.0   &     0.03   &	  1.23  &    1       \\   
    VCC415                  & -          & 12:20:25.30 & +06:54:32.0 & 0.0192   &  1.45   &   12.79    &     102    &     0.59  &   9.09    &      W     &    32.0  &	0.17  &     0.24   &	  0.8	&    40      \\   
    VCC428                  & -          & 12:20:40.35 & +13:53:20.3 & 0.0418   &  0.5    &   3.48     &     33     &     0.43  &   6.92    &      A     &    16.5  &	0.04  &     0.04   &	  0.54  &    11      \\
\hline
\end{tabular}}
\]
}
\end{table}
\end{landscape}

\begin{landscape}
\setcounter{table}{0}
\begin{table}
\caption{continued
}
\label{gal}
{\tiny 
\[
\resizebox{\columnwidth}{!}{\begin{tabular}{cccccccccccccccc}
\hline
\noalign{\smallskip}
\hline
Name		& NGC/IC  & RA(J2000)   & DEC(J2000)    & $E(B-V)$        &$2a_{B25.5}$ &$R_{eff,i}$&$PA(i)$    & $b/a$   &log $M_{star}$&SM     &dist        &$A(H\alpha)$&[{\nii}]/H$\alpha$&$HI-def$&N{\hii}$_{L(H\alpha)\geq 37}$\\
		&	  &  h m s      & $^o$ ' "      &                 & arcmin     & arcsec     & deg       &         & M$_{\odot}$  &       & Mpc	      & mag	   &	      &           &       \\
\hline
    VCC697                  & 3267       & 12:24:05.53 & +07:02:28.7 & 0.0234   &  1.45   &   22.89    &     47     &     0.8   &   9.48    &      W1    &    23.0  &	0.0   &     0.45   &	  1.02  &    56      \\   
    VCC740                  & -          & 12:24:39.61 & +08:30:13.3 & 0.0248   &  0.89   &   9.33     &     126    &     0.43  &   8.02    &      B     &    15.8  &	0.0   &     0.06   &	  0.62  &    6       \\   
    VCC758                  & 4370       & 12:24:54.91 & +07:26:40.0 & 0.0233   &  1.82   &   15.03    &     83     &     0.55  &   10.24   &      W1    &    23.0  &	0.68  &     0.63   &	  0.73  &    15      \\   
    VCC764                  & -          & 12:25:05.61 & +05:19:44.9 & 0.0323   &  0.54   &   4.46     &     125    &     0.69  &   8.69    &      B     &    15.8  &	0.0   &     0.28   &	  0.94  &    4       \\   
    VCC768                  & 3298       & 12:25:03.71 & +17:00:58.5 & 0.0265   &  1.29   &   8.66     &     161    &     0.36  &   8.53    &      A     &    16.5  &	0.13  &     0.23   &	  1.0	&    6       \\   
    VCC781                  & 3303       & 12:25:15.20 & +12:42:52.4 & 0.0373   &  1.17   &   9.96     &     67     &     0.65  &   8.86    &      A     &    16.5  &	0.0   &     0.22   &	  1.92  &    2       \\
\hline
\end{tabular}}
\]
}
\end{table}
\end{landscape}

\begin{landscape}
\setcounter{table}{0}
\begin{table}
\caption{continued
}
\label{gal}
{\tiny 
\[
\resizebox{\columnwidth}{!}{\begin{tabular}{cccccccccccccccc}
\hline
\noalign{\smallskip}
\hline
Name		& NGC/IC  & RA(J2000)   & DEC(J2000)    & $E(B-V)$        &$2a_{B25.5}$ &$R_{eff,i}$&$PA(i)$    & $b/a$   &log $M_{star}$&SM     &dist        &$A(H\alpha)$&[{\nii}]/H$\alpha$&$HI-def$&N{\hii}$_{L(H\alpha)\geq 37}$\\
		&	  &  h m s      & $^o$ ' "      &                 & arcmin     & arcsec     & deg       &         & M$_{\odot}$  &       & Mpc	      & mag	   &	      &           &       \\
\hline
    VCC1188                 & 796        & 12:29:26.32 & +16:24:17.5 & 0.031    &  1.95   &   11.22    &     138    &     0.38  &   9.34    &      A    &    16.5  &   0.75  &     0.32   &	 2.0   &    10      \\   
    VCC1189                 & 3414       & 12:29:29.00 & +06:46:12.7 & 0.0186   &  2.0    &   16.1     &     20     &     0.36  &   8.94    &      B    &    15.8  &   0.41  &     0.28   &	 0.83  &    66      \\   
    VCC1190                 & 4469       & 12:29:28.02 & +08:44:59.2 & 0.0201   &  5.37   &   23.12    &     48     &     0.53  &   10.29   &      B    &    15.8  &   1.0   &     0.77   &	 2.99  &    15      \\   
    VCC1193                 & 4466       & 12:29:30.55 & +07:41:47.0 & 0.0257   &  1.51   &   11.57    &     100    &     0.37  &   8.91    &      B    &    15.8  &   0.51  &     0.35   &	 0.74  &    64      \\   
    VCC1200                 & 3416       & 12:29:35.58 & +10:47:37.1 & 0.0314   &  1.07   &   15.16    &     69     &     0.5   &   8.36    &      A    &    16.5  &   0.0   &     0.09   &	 2.03  &    12      \\   
    VCC1205                 & 4470       & 12:29:37.79 & +07:49:26.8 & 0.0241   &  2.0    &   13.69    &     19     &     0.55  &   9.32    &      B    &    15.8  &   1.05  &     0.44   &	 0.58  &    148     \\
\hline
\end{tabular}}
\]
}
\end{table}
\end{landscape}

\begin{landscape}
\setcounter{table}{0}
\begin{table}
\caption{continued
}
\label{gal}
{\tiny 
\[
\resizebox{\columnwidth}{!}{\begin{tabular}{cccccccccccccccc}
\hline
\noalign{\smallskip}
\hline
Name		& NGC/IC  & RA(J2000)   & DEC(J2000)    & $E(B-V)$        &$2a_{B25.5}$ &$R_{eff,i}$&$PA(i)$    & $b/a$   &log $M_{star}$&SM     &dist        &$A(H\alpha)$&[{\nii}]/H$\alpha$&$HI-def$&N{\hii}$_{L(H\alpha)\geq 37}$\\
		&	  &  h m s      & $^o$ ' "      &                 & arcmin     & arcsec     & deg       &         & M$_{\odot}$  &       & Mpc	      & mag	   &	      &           &       \\
\hline
    VCC1690                 & 4569       & 12:36:49.86 & +13:09:46.7 & 0.0464   &  13.18  &   72.06    &     16     &     0.55  &   10.74   &      A     &    16.5  &	2.82  &     0.97   &	  1.07  &    691     \\   
    VCC1696                 & 4571       & 12:36:56.39 & +14:13:02.4 & 0.0461   &  4.68   &   49.53    &     63     &     0.9   &   10.18   &      A     &    16.5  &	0.45  &     0.52   &	  0.73  &    365     \\   
    VCC1699                 & 3591       & 12:37:02.81 & +06:55:34.7 & 0.0235   &  1.45   &   15.14    &     49     &     0.31  &   8.65    &      B     &    15.8  &	0.23  &     0.15   &	  0.53  &    41      \\   
    VCC1715                 & -          & 12:37:28.47 & +08:47:40.0 & 0.0221   &  0.36   &   7.76     &     31     &     0.88  &   8.1     &      B     &    15.8  &	0.0   &     0.07   &	  1.23  &    2       \\   
    VCC1725                 & -          & 12:37:41.14 & +08:33:33.1 & 0.0226   &  1.45   &   20.18    &     116    &     0.39  &   8.71    &      B     &    15.8  &	0.06  &     0.19   &	  1.05  &    62      \\   
    VCC1726                 & 7795       & 12:37:45.42 & +07:06:21.4 & 0.02     &  1.62   &   20.8     &     16     &     0.76  &   7.98    &      B     &    15.8  &	0.0   &     0.09   &	  0.41  &    56      \\
\hline
\end{tabular}}
\]
}
\end{table}
\end{landscape}

\begin{landscape}
\setcounter{table}{0}
\begin{table}
\caption{continued
}
\label{gal}
{\tiny 
\[
\resizebox{\columnwidth}{!}{\begin{tabular}{cccccccccccccccc}
\hline
\noalign{\smallskip}
\hline
Name		& NGC/IC  & RA(J2000)   & DEC(J2000)    & $E(B-V)$        &$2a_{B25.5}$ &$R_{eff,i}$&$PA(i)$    & $b/a$   &log $M_{star}$&SM     &dist        &$A(H\alpha)$&[{\nii}]/H$\alpha$&$HI-def$&N{\hii}$_{L(H\alpha)\geq 37}$\\
		&	  &  h m s      & $^o$ ' "      &                 & arcmin     & arcsec     & deg       &         & M$_{\odot}$  &       & Mpc	      & mag	   &	      &           &       \\
\hline
    AGC226131               & -          & 12:21:13.25 & +10:37:32.7 & 0.032    &  0.52   &   5.39     &     14     &     0.64  &   6.63    &      A     &    16.5  &	0.0   &     0.01   &	  0.55  &    2       \\   
    AGC227861               & -          & 12:29:59.50 & +08:25:54.5 & 0.0206   &  0.7    &   7.29     &     179    &     0.83  &   8.19    &      B     &    15.8  &	0.0   &     0.1    &	  0.64  &    1       \\   
    AGC224516               & -          & 12:35:24.39 & +05:02:53.5 & 0.0236   &  0.58   &   6.02     &     105    &     0.8   &   7.54    &      B     &    15.8  &	0.12  &     0.09   &	  0.62  &    5       \\   
    AGC229323               & -          & 12:44:51.86 & +16:25:06.2 & 0.0304   &  0.78   &   8.06     &     66     &     0.59  &   7.59    &      A     &    16.5  &	0.0   &     0.07   &	  0.41  &    11      \\   
    IC3806                  & -          & 12:48:55.36 & +14:54:28.4 & 0.0284   &  1.29   &   13.37    &     28     &     0.52  &   9.24    &      A     &    16.5  &	0.0   &     0.39   &	  2.01  &    45      \\   
    N12:10:32.04+14:29:23.3 & -          & 12:10:32.04 & +14:29:23.3 & 0.0344   &  0.63   &   6.49     &     101    &     0.78  &   8.18    &      M     &    32.0  &	0.0   &     0.2    &	  0.95  &    5       \\
\hline
\end{tabular}}
\]
Column 1: galaxy name \\
Column 2: IC/NGC name \\
Column 3 and 4: right ascension and declination\\
Column 5: Galactic extinction $E(B-V)$, from Schlegel et al. (1998)\\
Column 6: $B$-band isophotal diameter at 25.5 mag~arcsec$^{-2}$, from Binggeli et al. (1985) for all the VCC galaxies, from GoldMine (Gavazzi et al. 2003) for the NGC galaxies, 
or derived from the NGVS $g$-band effective radius using the relation $2 \times a_{B25.5}$ [arcsec] = 3.9886 $\times$ $R_{eff,g}$ [arcsec] for the remaining objects \\
Column 7: NGVS $i$-band effective radius $R_{eff,i}$ \\
Column 8: NGVS $i$-band position angle, measured from North counterclockwise\\
Column 9: $B$-band axial ratio, from Binggeli et al. (1985) whenever available, or from the NGVS $g$-band\\
Column 10: stellar mass, in solar units\\
Column 11: cluster subgroup membership\\
Column 12: distance, in Mpc\\
Column 13: $A(H\alpha)$, mag\\
Column 14: [{\nii}]$\lambda$6548,6583/H$\alpha$\\
Column 15: $HI-def$\\
Column 16: number of {\hii} regions brighter than $L(H\alpha)$$\geq$ 10$^{37}$ erg~s$^{-1}$ (luminosity corrected for dust attenuation and {\nii} contamination)\\
}
\end{table}
\end{landscape}
\newpage

\begin{table*}[h!]
\caption{Apertures used for the flux extraction and for the identification of the {\hii} regions
}
\label{Tabds9}
{\tiny 
\[
\begin{tabular}{cccccc}
\hline
\noalign{\smallskip}
\hline
Name	    & RA(J2000)   & DEC(J2000)  &Maj. axis& Min. axis   & PA   \\
	    & h m s	  & $^o$ ' "	& arcsec  & arcsec      & deg. \\
\hline
   VCC15   	     & 12:09:53.48 & +13:03:01.6 & 33.89   &	    23.56    &       0       \\ 
   VCC34   	     & 12:11:09.97 & +13:35:15.6 & 32.78   &	    20.63    &       0       \\ 
   VCC41   	     & 12:12:03.36 & +12:44:31.9 & 14.39   &	    4.671    &       177     \\ 
   VCC47   	     & 12:12:11.20 & +13:14:47.3 & 37.05   &	    20.28    &       164     \\ 
   VCC48   	     & 12:12:17.18 & +12:29:07.0 & 65.43   &	    46.36    &       45      \\ 
   VCC67   	     & 12:12:45.05 & +13:58:35.8 & 64.2    &	    23.71    &       67      \\
\hline
\end{tabular}
\]
}
\end{table*}
\newpage

\setcounter{table}{1}
\begin{table*}
\caption{continued
}
\label{Tabds9}
{\tiny
\[
\begin{tabular}{cccccc}
\hline
\noalign{\smallskip}
\hline
Name	    & RA(J2000)   & DEC(J2000)  &Maj. axis& Min. axis   & PA   \\
	    & h m s	  & $^o$ ' "	& arcsec  & arcsec      & deg. \\
\hline
  VCC620    	    & 12:23:09.50 & +11:43:34.5 & 37.3    &	   12.07    &	    147     \\ 
  VCC630    	    & 12:23:01.04 & +11:21:34.0 & 125.4   &	   33.94    &	    57      \\ 
  VCC641    	    & 12:23:25.85 & +05:49:00.3 & 15.27   &	   7.896    &	    100     \\ 
  VCC656    	    & 12:23:38.64 & +06:57:14.9 & 80.42   &	   33.35    &	    130     \\ 
  VCC664    	    & 12:23:44.40 & +12:28:41.3 & 79.56   &	   52.09    &	    95      \\ 
  VCC667    	    & 12:23:48.49 & +07:11:12.5 & 51.1    &	   26.42    &	    17      \\
\hline
\end{tabular}
\]
}
\end{table*}
\newpage

\setcounter{table}{1}
\begin{table*}
\caption{continued
}
\label{Tabds9}
{\tiny
\[
\begin{tabular}{cccccc}
\hline
\noalign{\smallskip}
\hline
Name	    & RA(J2000)   & DEC(J2000)  &Maj. axis& Min. axis   & PA   \\
	    & h m s	  & $^o$ ' "	& arcsec  & arcsec      & deg. \\
\hline
     VCC1486  	        & 12:33:10.09 & +11:20:50.6 & 17.15   &	   13.03    &	    170     \\ 
     VCC1501  	        & 12:33:23.92 & +08:41:24.2 & 14.99   &	   6.687    &	    30      \\ 
     VCC1516  	        & 12:33:31.41 & +09:10:47.6 & 82.72   &	   45.43    &	    30      \\ 
     VCC1532  	        & 12:33:56.66 & +15:21:16.8 & 41.77   &	   26.39    &	    160     \\ 
     VCC1535  	        & 12:34:03.00 & +07:41:57.8 & 18.09   &	   6.331    &	    110     \\ 
     VCC1552  	        & 12:34:15.72 & +13:04:31.3 & 35.13   &	   26.96    &	    155     \\
\hline
\end{tabular}
\]
}
\end{table*}
\newpage

\setcounter{table}{1}
\begin{table*}
\caption{continued
}
\label{Tabds9}
{\tiny
\[
\begin{tabular}{cccccc}
\hline
\noalign{\smallskip}
\hline
Name	    & RA(J2000)   & DEC(J2000)  &Maj. axis& Min. axis   & PA   \\
	    & h m s	  & $^o$ ' "	& arcsec  & arcsec      & deg. \\
\hline
N12:14:43.05+10:31:59.3 & 12:14:43.13 & +10:31:58.7 & 4.574   &        2.655	&	0	\\ 
N12:16:12.71+07:27:21.6 & 12:16:13.73 & +07:27:21.6 & 4.488   &        4.488	&	90	\\ 
N12:16:20.63+14:46:26.9 & 12:16:21.22 & +14:46:28.1 & 6.006   &        6.006	&	90	\\ 
N12:18:45.13+06:15:47.4 & 12:18:45.42 & +06:15:48.9 & 2.036   &        2.036	&	90	\\ 
N12:19:28.67+17:13:49.8 & 12:19:28.47 & +17:13:49.7 & 4.85    &        4.85	&	90	\\ 
N12:20:20.16+09:47:49.5 & 12:20:19.89 & +09:47:49.2 & 7.829   &        7.829	&	90	\\
\hline
\end{tabular}
\]
Column 1: galaxy name \\
Column 2 and 3: right ascension and declination of the centre of the elliptical aperture \\
Column 4 and 5: semi major and minor axis of the elliptical aperture, in arcsec \\
Column 6: position angle of the elliptical aperture, measured from north counter clockwise \\ 
}
\end{table*}
\newpage

\begin{table*}
\caption{Best-fit parameters of the fit of the luminosity function for individual galaxies
}
\label{galLF}
{\tiny
\[
\begin{tabular}{cccccccccc}
\hline
\noalign{\smallskip}
\hline
Name		  &$\alpha_{16}$& $\alpha$&$\alpha_{84}$&log$L^*(H\alpha)_{16}$&log$L^*(H\alpha)$&log$L^*(H\alpha)_{84}$&$\Phi^*_{16}$&$\Phi^*$&$\Phi^*_{84}$	  \\
		  &	   &	     &  	   & erg~s$^{-1}$ & erg~s$^{-1}$ & erg~s$^{-1}$ & N dex$^{-1}$ & N dex$^{-1}$ & N dex$^{-1}$ \\
\hline
       VCC15  &    -4.34  &    -3.72   &   -3.16  &    38.45    &  39.53  &    40.52  &       -   &    -      &    -    \\
       VCC47  &    -1.68  &    -1.55   &   -1.41  &    38.73    &  39.01  &    39.47  &     0.38  &     0.85  &     1.17\\
       VCC48  &    -2.05  &    -1.83   &   -1.60  &    38.99    &  39.75  &    40.58  &    -1.82  &    -0.93  &    -0.05\\
       VCC67  &    -2.05  &    -1.84   &   -1.63  &    39.08    &  39.81  &    40.59  &    -1.89  &    -0.96  &    -0.15\\
       VCC83  &    -2.18  &    -1.95   &   -1.70  &    38.58    &  39.36  &    40.42  &    -1.96  &    -0.68  &     0.41\\
       VCC97  &    -1.76  &    -1.71   &   -1.65  &    39.46    &  39.70  &    40.09  &     0.35  &     0.74  &     0.99\\
\hline
\end{tabular}
\]
}
\end{table*}
\newpage
\clearpage

\setcounter{table}{2}
\begin{table*}[h!]
\caption{continued
}
\label{galLF}
{\tiny
\[
\begin{tabular}{cccccccccc}
\hline
\noalign{\smallskip}
\hline
Name		  &$\alpha_{16}$& $\alpha$&$\alpha_{84}$&log$L^*(H\alpha)_{16}$&log$L^*(H\alpha)$&log$L^*(H\alpha)_{84}$&$\Phi^*_{16}$&$\Phi^*$&$\Phi^*_{84}$	  \\
		  &	   &	     &  	   & erg~s$^{-1}$ & erg~s$^{-1}$ & erg~s$^{-1}$ & N dex$^{-1}$ & N dex$^{-1}$ & N dex$^{-1}$ \\
\hline
     VCC1686  &    -1.60  &    -1.48   &   -1.36  &    38.97    &  39.23  &    39.68  &     0.33  &     0.78  &     1.07\\
     VCC1690  &    -1.60  &    -1.57   &   -1.53  &    39.82    &  39.99  &    40.22  &     0.77  &     0.98  &     1.13\\
     VCC1696  &    -2.28  &    -2.19   &   -2.07  &    38.71    &  39.12  &    40.10  &    -1.09  &     0.23  &     0.89\\
     VCC1699  &    -1.88  &    -1.75   &   -1.64  &    40.11    &  40.52  &    40.84  &    -1.52  &    -1.06  &    -0.59\\
     VCC1725  &    -1.95  &    -1.77   &   -1.55  &    38.56    &  39.03  &    40.14  &    -1.02  &     0.20  &     0.83\\
     VCC1726  &    -2.22  &    -2.01   &   -1.75  &    38.46    &  39.22  &    40.36  &    -1.85  &    -0.42  &     0.68\\
\hline
\end{tabular}
\]
Column 1: galaxy name \\
Column 2-4: $\alpha$ parameter of the best-fit Schechter function with uncertainty estimates as the 16th and 84th percentiles
of the marginalised posterior distribution\\
Colunn 5-7: $L^*(H\alpha)$ parameter of the best-fit Schechter function with uncertainty estimates as the 16th and 84th percentiles
of the marginalised posterior distribution\\
Column 8-10: $\Phi^*$ parameter of the best-fit Schechter function with uncertainty estimates as the 16th and 84th percentiles
of the marginalised posterior distribution\\
}
\end{table*}
\newpage
\clearpage

\section{Luminosity function on individual objects}

Figure \ref{LFindividual} shows the luminosity function of \hii\ regions derived for the 100 galaxies of the sample
with at least 20 \hii\ regions detected by \textsc{HIIphot} above the completeness limitr of 
$L(H\alpha)$ $\geq$ 10$^{37}$ erg~s$^{-1}$. This figure can be compared to Fig. E.1 in Boselli et al. (2025) 
for the 27 unperturbed galaxies.

\begin{figure*}[h!]
\centering
\includegraphics[width=0.99\textwidth]{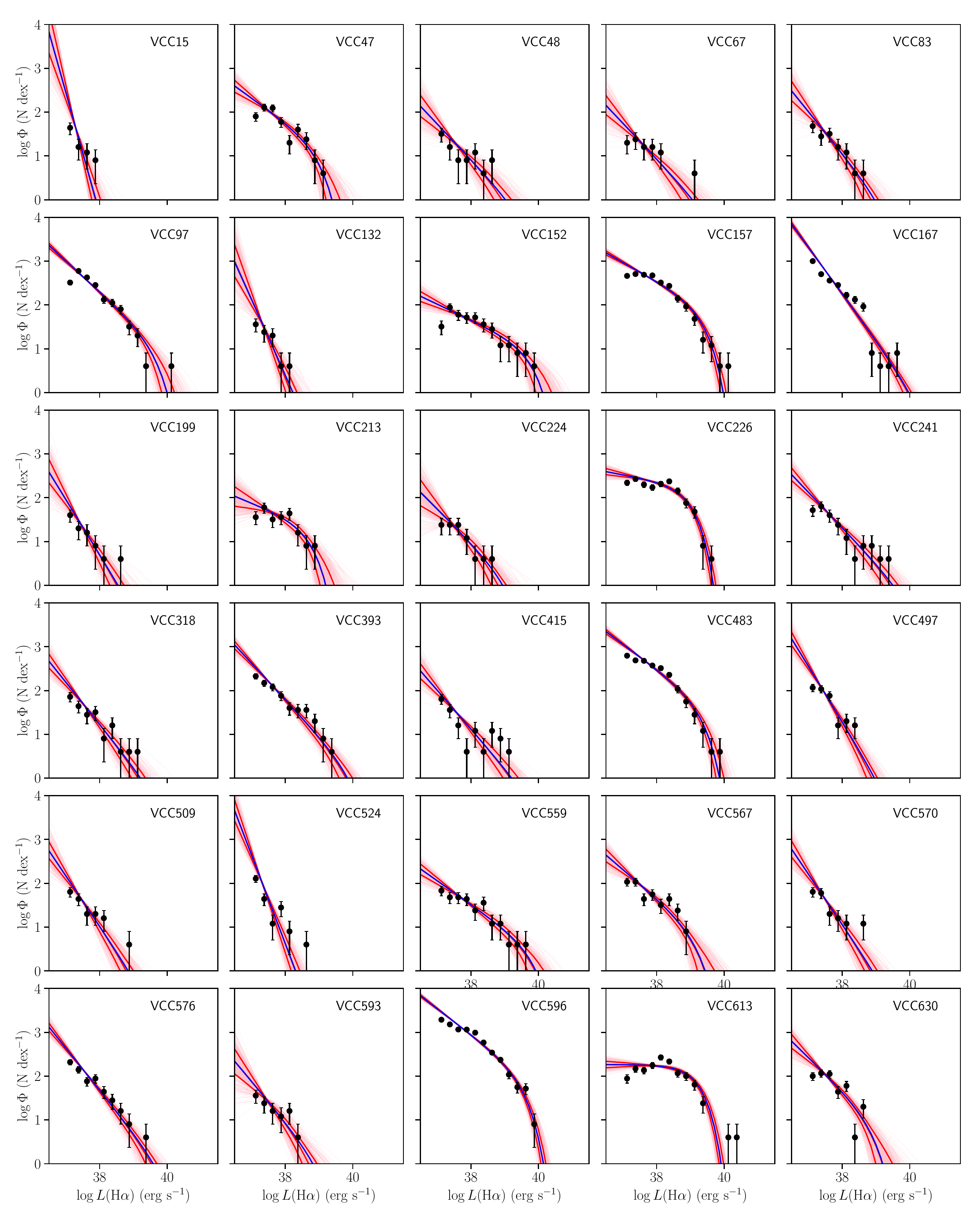}\\
\caption{Luminosity function of the {\hii} regions detected by \textsc{HIIphot} in individual galaxies. The H$\alpha$ luminosity of 
individual {\hii} regions is corrected for dust attenuation and [{\nii}] contamination as described in Sec. 3.2. The solid  blue and red lines
indicate the best fit and 1$\sigma$ confidence regions for the Schechter luminosity function parametrisation. Black solid dots indicate the number of {\hii} regions
in 0.25 dex bins of H$\alpha$ luminosity above the adopted completeness of the survey ($L(H\alpha)$ $\geq$ 10$^{37}$ erg~s$^{-1}$).}
\label{LFindividual}%
\end{figure*}

\setcounter{figure}{0}
\begin{figure*}[h!]
\centering
\includegraphics[width=0.99\textwidth]{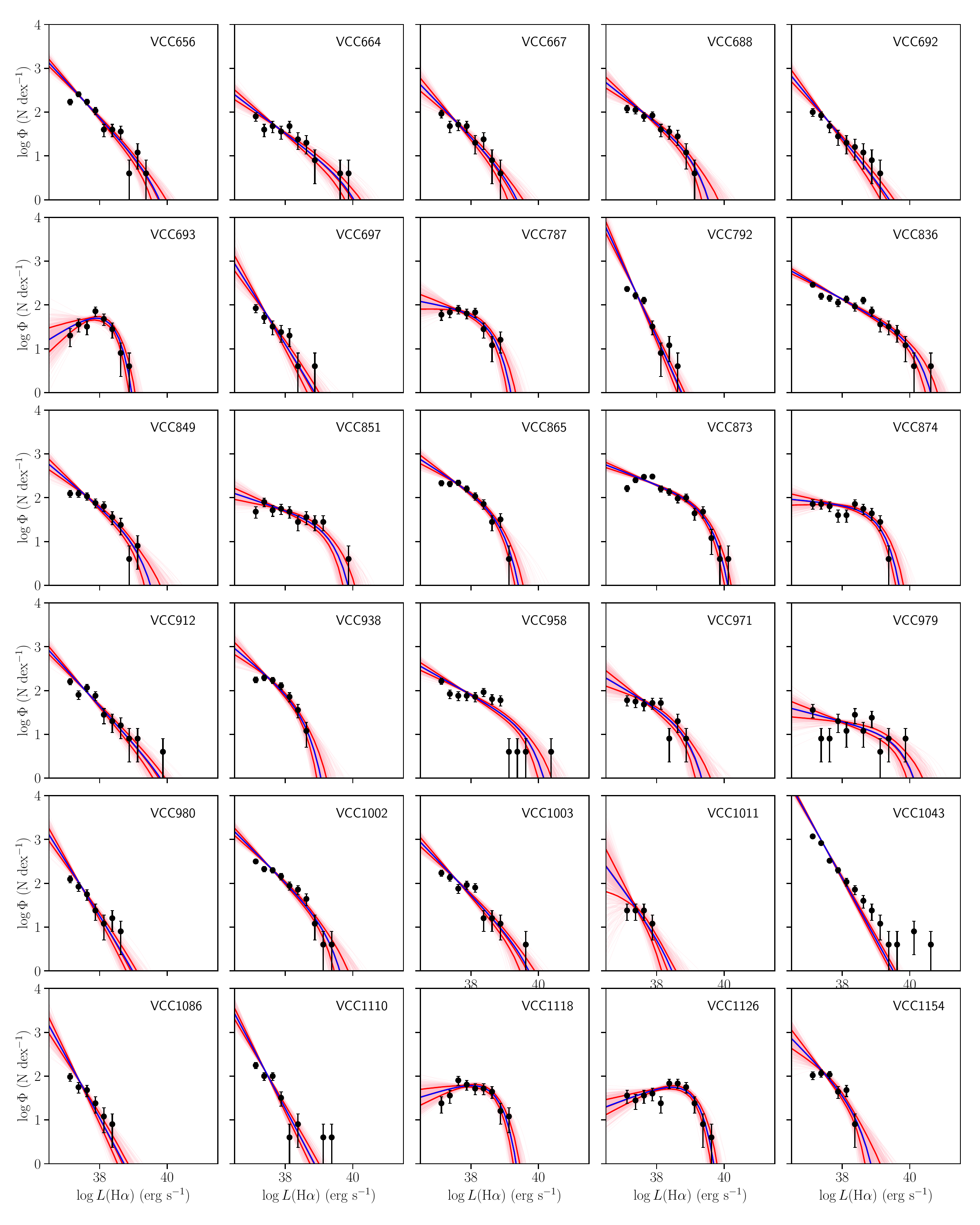}\\
\caption{Continued}
\label{LFindividual}%
\end{figure*}

\setcounter{figure}{0}
\begin{figure*}[h!]
\centering
\includegraphics[width=0.99\textwidth]{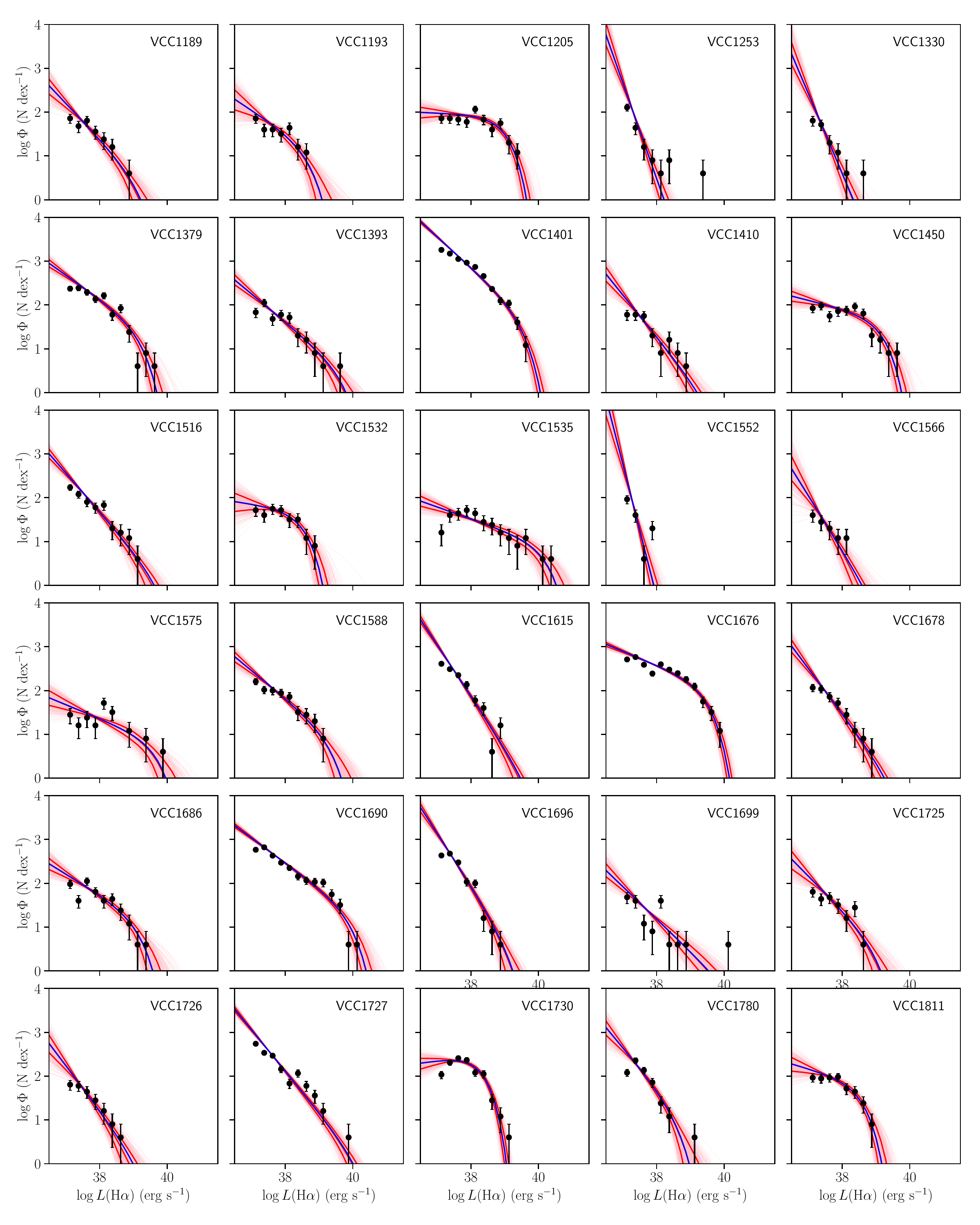}\\
\caption{Continued}
\label{LFindividual}%
\end{figure*}

\setcounter{figure}{0}
\begin{figure*}[h!]
\centering
\includegraphics[width=0.99\textwidth]{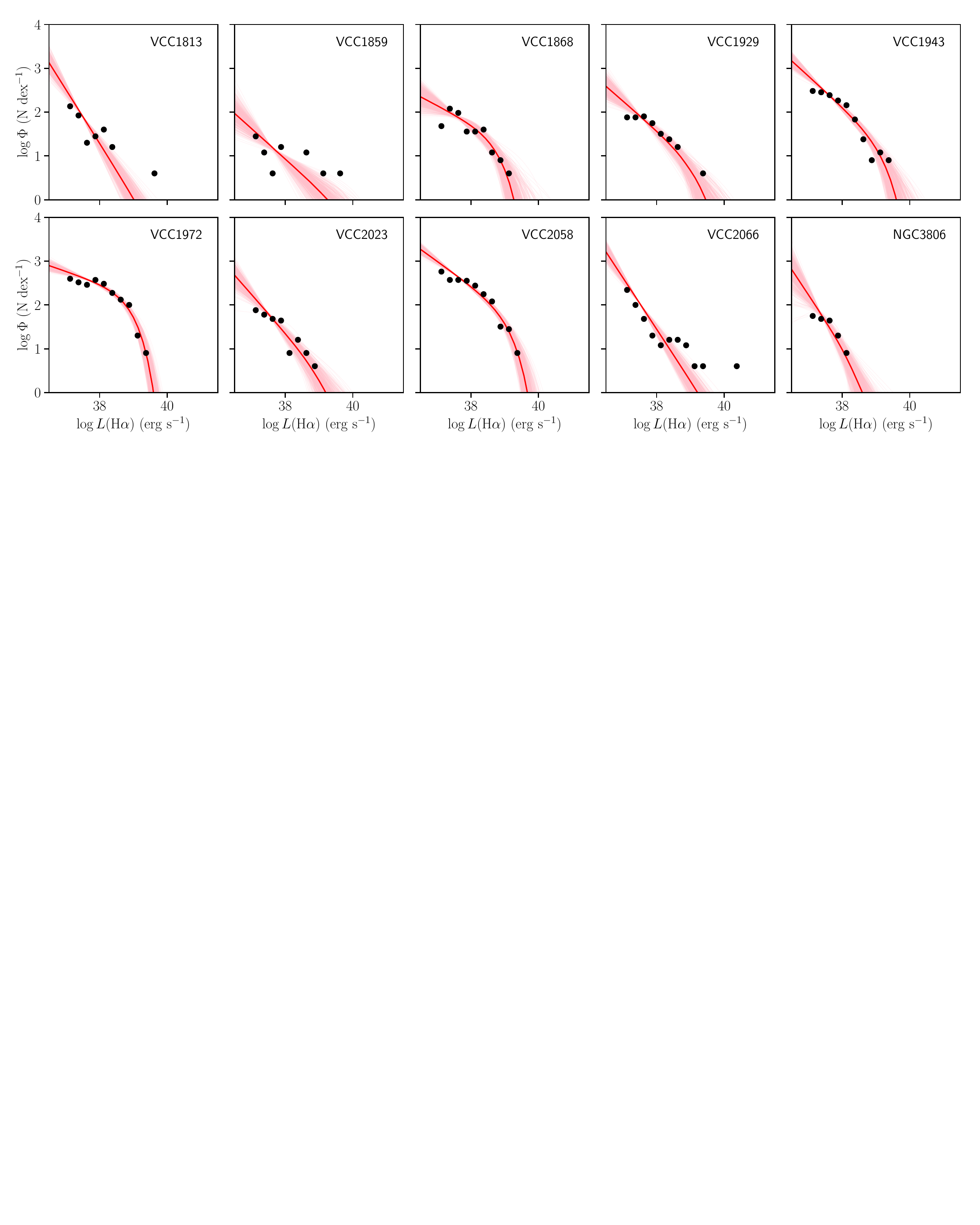}\\
\caption{Continued}
\label{LFindividual}%
\end{figure*}

\end{appendix}
\end{document}